\documentclass[twoside,a4paper]{article}

\def\versionName{Equivalence of Correlations}

\usepackage{oldgerm,amsmath,pifont,amsfonts,amssymb,latexsym}

\usepackage{lmodern}

\usepackage{graphicx}
\usepackage{subfig}
\usepackage{epstopdf}
\usepackage{float}

\usepackage{mathrsfs} 
\usepackage[latin1]{inputenc} 
\usepackage[T1]{fontenc}
\usepackage[english]{babel}
\usepackage{graphicx}

\usepackage{xargs}

\usepackage{natbib}
\renewcommand{\cite}[1]{\citep{#1}}

\usepackage{fancyhdr}
\usepackage{hyperref}
\usepackage[normalem]{ulem} %cross line of text

\usepackage{amssymb}

\newcommand {\eq}  [1] {\eqref{eq:#1}}
\newcommand {\Eq}  [1] {Eq.~\eq{#1}}
\newcommand {\Eqs} [1] {Eqs.~\eq{#1}}
\newcommand {\EQ}  [1] {Equation~\eq{#1}}

\newcommand {\Lemma} [1] {Lemma~\ref{lem:#1}}
\newcommand {\Propo} [1] {Proposition~\ref{prop:#1}}

\newcommand {\ie} {\emph{i.e.}}
\newcommand {\cf} {\emph{cf.}}

\newcommand{\za}{\alpha}
\newcommand{\zb}{\beta}
\newcommand{\zg}{\gamma}

\newcommand{\zmsd}{\ensuremath{\zg}}
\newcommand{\ah}{\`a\ }

\newcommand{\hX}{\hat{\tt x}}
\newcommand{\hY}{\hat{\tt y}}
\newcommand{\bm}{\overline{m}} 
\newcommand{\bt}[1]{\ensuremath{{\widetilde{m}_{\za, #1}(x)}}}

\newcommand{\N}{\mathbb{N}} 
\newcommand{\Z}{\mathbb{Z}} 
\newcommand{\R}{\mathbb{R}} 
\newcommand{\llgp}{LLg$^+$}
\newcommand{\Eo}{\mathbb{E}_\omega}
\newcommand{\Ey}{\mathbb{E}_{{\tt Y}}}
\newcommand{\E}{\mathbb{E}}
\newtheorem{thm}{Theorem}
\newtheorem{lem}[thm]{Lemma}
\newtheorem{prop}[thm]{Proposition}
\newtheorem{conj}[thm]{Conjecture}
\newtheorem{oss}[thm]{Remark}
\newcommand{\proof}{\emph{Proof.}\, }
\newcommand{\evidence}{\emph{Numerical Evidence.}\, }

\newcommand{\keywords}[1]{\qquad\, \textbf{Keywords:}~#1}

\title{Equivalence of position-position auto-correlations\\ in the Slicer Map and the L\'evy-Lorentz gas}

\author{C. Giberti$^{\textup{{\tiny(a)}}}$,
  L. Rondoni$^{\textup{{\tiny (b,c)}}}$,
  M. Tayyab$^{\textup{{\tiny(b,d)}}}$ and
  J. Vollmer$^{\textup{{\tiny(b,e)}}}$
\\
{\small $^{\textup{(a)}}$
 Dipartimento di Scienze e Metodi dell'Ingegneria,
Universit\`a di Modena e Reggio E., }
\\
{\small  Via Amendola 2, Padiglione Morselli, I-42122 Reggio E., Italy}
\\
{\small $^{\textup{(b)}}$ Dipartimento di Scienze Matematiche, Giuseppe Luigi Lagrange, Politecnico di Torino,
}\\
{\small Corso Duca degli Abruzzi 24 I-10129 Torino, Italy}
\\
{\small $^{\textup{(c)}}$ Present address: Civil and Environmental Engineering Department, and Princeton Environmental Institute,}
\\
{\small Princeton University, 59 Olden St, Princeton, NJ 08540, USA}
\\
{\small $^{\textup{(d)}}$ Dipartimento di Matematica, Giuseppe Peano, Universit\`{a} degli Studi di Torino,}
\\
{\small Via Carlo Alberto 10, I-10123 Torino, Italy}
\\
{\small $^{\textup{(e)}}$ Present address: Institut f\"ur Theoretische Physik, Universit\"at Leipzig, Br\"uderstr.~16, D-04103 Leipzig, Germany}
}

\date{\today}

\begin{document}

\maketitle

\begin{abstract}
The Slicer Map is a one-dimensional non-chaotic dynamical system that shows sub-, super-, 
and normal diffusion as a function of its control parameter. 
In a recent paper [Salari et al., CHAOS 25, 073113 (2015)] it was found
that the moments of the position distributions as the Slicer Map have the same asymptotic behaviour as the L\'evy-Lorentz gas,
a  random walk on the line in which the scatterers are randomly distributed  according  to a L\'evy-stable probability distribution.
Here we derive analytic expressions for the position-position correlations of the Slicer Map and,
on the ground of  this result, we formulate some conjectures
about the  asymptotic behaviour of position-position correlations of the L\'evy-Lorentz gas,
for which the information in the literature is minimal. 
The numerically estimated position-position  correlations of the L\'evy-Lorentz show a remarkable agreement 
with the conjectured asymptotic scaling.
\end{abstract}
\keywords{Slicer Map, L\'evy-Lorentz lattice gas, position-position auto-correlation function, anomalous transport.}

%%%%  ------------------------------------------------------------------------------------------   %%%%
%%%%  ------------------------------------------------------------------------------------------   %%%%
\section{Introduction}
\label{sec:Introduction}

One object of interest in studies of anomalous transport is the \emph{transport exponent} $\gamma$
\cite{KRS08}:
\begin{equation}\label{eq:gamma}
  \zg := \lim_{n \to \infty} \frac{\log \langle \Delta {\bf x}^{2}_{n}\rangle}{\log n} \, ,
\end{equation}
where $\langle \Delta {\bf x}^{2}_{n}\rangle$ is the mean-square displacement of the positions 
at time $n$.
Regimes with $\zg < 1$ are called sub-diffusion; they are called diffusion if $\zg=1$, and 
super-diffusion if $\zg > 1$.
Some special cases treated in \citet{Salari}, such as logarithmic growth of 
$\langle \Delta {\bf x}^{2}_{n}\rangle$, are not 
described by \Eq{gamma}, which merely yields $\gamma=0$, \cf~Theorem \ref{thm:teorema1}.
However, when \Eq{gamma} holds  with $\gamma >0$ the generalised diffusion coefficient $D_\zg$, defined by:
\begin{equation}
  D_\zg := \lim_{n\to\infty} \frac{\langle \Delta {\bf x}^{2}_{n}\rangle}{n^\zg} \, ,
\end{equation}
exists and is  a non-negative number.

Transport properties afford only a rather coarse representation of the typically very 
rich underlying microscopic dynamics.
Understanding them from a microscopic perspective is an open problem,
that motivates a wide and very active research community \cite{KlLiMe,Kla06,VulpTrans,KRS08,JBS03,Sokolov}.
In the realm of deterministic dynamics, it is understood that uniformly hyperbolic dynamical 
systems produce rapid correlations decay. In turn, rapid 
decay of correlations is commonly associated with standard diffusion \cite{Gas05,Kla06}. 
Because randomly placed non-overlapping wind-trees%
\footnote{In wind-tree models a ``wind'' particle moves
  with constant velocity on a plane,
  where it is elastically reflected at fixed square scatterers, the ``trees''. }
and related maps \cite{DC00,CFVN02} enjoy a sort of stochasticity analogous to that generated by chaotic
dynamics, they may also show standard diffusion.  

On the contrary, for fully deterministic systems with vanishing Lyapunov exponents, such as 
systems of point particles within periodic polygonal walls \cite{JeRo06}, the nature of transport is still a matter of 
investigation \cite{Zas02,Kla06,KRS08,Salari}. 
A major challenge of the latter systems is that correlations persist or decay rather slowly as
compared to what happens in chaotic systems \cite{JeRo06}.
This makes their asymptotic statistics much harder to understand than in the presence of chaos.
Indeed, the unpredictability of single trajectories in
strongly chaotic systems, like axiom A systems, that is one aspect of the 
fast decay of correlations, is often associated with regular behaviour on the level of ensembles, as 
proven, for instance, by the differentiability of SRB states \cite{Rue97} that implies linear response \cite{Rue98}. 
In contrast,  for non-chaotic systems the parameter dependence of the transport exponent
can be quite irregular \cite{JeRo06}.
(Generalized) diffusion coefficients may be irregular even in chaotic systems \cite{Kla06}.

In the field of fully fledged stochastic processes, numerous questions remain open as well
\cite{Zas02,Kla06,DKU03,LWWZ05,Sokolov}. 
Among such systems, the L\'evy-Lorentz gas (LLg), a random walk in random environments, 
in which the scatterers are randomly distributed on a line
according to a L\'evy-stable probability distribution, 
has been thoroughly investigated by various authors.
Different types of anomalous and standard diffusion were observed,
upon tuning the parameter~$\zb$ 
characterising the  L\'evy-stable probability distribution \cite{BF99,BFK00}.
These authors noted that ballistic contributions to the mean-square displacement,
which are considered irrelevant when diffusion is normal,
are in fact important for the transport.
Under certain simplifying assumptions, \citet{BCV10}  analytically calculated 
the mean-square displacement of the travelled distance for this model, 
and numerically verified the validity of their reasoning.
More recently, \citet{BCLL} rigorously established the validity of the Central Limit Theorem.

The Slicer Map (SM) introduced in \citet{Salari} was motivated by observations of the  
mass transport of periodic polygonal billiards \cite{JeRo06}.
Like in polygonal billiards, the dynamics of the SM are free of randomness.
Their trajectories do not separate exponentially in time,
and they experience sudden deviations from their motion, at 
isolated points that are regularly placed in space. 
Despite these facts, the dynamics of the SM differs substantially from all other models mentioned so far.
For instance, after an initial transient all trajectories of the SM turn periodic.
However, anomalous transport may be dominated by ballistic flights \cite{AKB}.
Indeed the SM features anomalous transport because
the length of ballistic flights in the initial ensemble follows a power-law distribution.
Its transport exponent $\gamma_\alpha$ can be tuned by adjusting its parameter~$\za$
that governs the power-law distribution of the ballistic flights.
 
\citet{Salari} showed that once $\za$ is adjusted so that the transport exponent of the SM coincides
with that of the LLg at a given $\zb$, 
all higher order moments of the position distribution of the SM scale in time like those of the LLg
\cite{BCV10}.
Of course such an agreement does not imply a full equivalence of the dynamics, as mentioned above and further 
stressed in Sec.~\ref{sec:slicer} and~\ref{sec:discussion}.
For instance, from the particle-transport viewpoint, the LLg can only be super-diffusive 
($1 < \zg_\zb < 2$), while  the~SM can exhibit all possible diffusion regimes, $0 \leq \zg_\za \leq 2$.
Moreover, the $\zb$-dependence of the transport exponent of the LLg is not simple: 
it splits in three different functional forms. In contrast, $\zg_\alpha = 2-\za$ for all SM regimes.

The agreement of the moments
of the displacement suggest an equivalence of the transport
characteristics of
the SM and the LLg.
This is similar to findings in statistical physics, 
where systems with different microscopic dynamics can also share the same thermodynamic properties
(\ie\ averages and variances) and corresponding correlation functions.
Nevertheless, such a correspondence is far from trivial.
In particular for a transient, far-from equilibrium dynamics with non-normal diffusion
a direct investigation is indispensible. 
The common wisdom is that it should be possible to identify differences in 
\emph{some} correlation functions \cite{KRS08,Sokolov}.

Here, we derive analytic expressions for the position-position
auto-correlations of the SM,
and we explore
whether they suffice to distinguish its transport properties from those of
the LLg and a closely related systems, the modified L\'evy-Lorentz gas (\llgp), that will be introduced in Sec.~\ref{sec:compareLW}.
As information on the position correlations of the LLg
and the \llgp\ is minimal in the literature,
we resort to numerical simulations
to compare the SM with the  LLg and the \llgp.
We find that the equivalence of the positions moments extends to the case of the position-position auto-correlation functions;
for $1.5 \lesssim \zg < 2$ their functional forms have been numerically found to match without adjustable parameters. 
Therefore, even these auto-correlations do not distinguish the Slicer Map and the (modified) L\'evy-Lorentz gas.

This paper is organised as follows: 
Section~\ref{sec:slicer} formally introduces the SM and summarises its main properties.
They are derived here in an alternative fashion, compared to that of \citet{Salari}. 
Some examples of the SM position auto-correlation functions are explicitly computed in Sec.~\ref{sec:corrslicer}.
The correspondence between the \llgp\ and the SM is discussed in Sec.~\ref{sec:compareLW}.
The position-position auto-correlations of the SM and the \llgp\ are compared in Sec.~\ref{sec:CorrLL}. 
More precisely, leveraging on the knowledge of  position-position auto-correlations of the SM, we propose some conjectures for the asymptotic scaling of the  correlations of the \llgp\ and  provide numerical evidence supporting them.
In Sec.~\ref{sec:discussion} we conclude the paper with a discussion of our main result:
In the strongly super-diffusive regime, $1.5 \lesssim \zg < 2$,
the position-position auto-correlations of the SM and of the \llgp\ scale in the same fashion with time. 
We interpret this finding in terms of the distribution of the length of ballistic flights,
which do not strongly depend on the details of the dynamics. 
For $1 \lesssim \zg < 1.5$, good statistic is harder to obtain; we expect the equivalence to
hold also in this parameter range, but at the moment we cannot properly support this expectation.
Some technical points of the proof, that concern the time asymptotics of the moments and 
the auto-correlation function, are provided in an appendix.

%%%%  ------------------------------------------------------------------------------------------   %%%%
%%%%  ------------------------------------------------------------------------------------------   %%%%
\section{The Slicer Map}
\label{sec:slicer}

To define the SM we introduce the fundamental space unit $M:=[0,1]$, consisting of the interval of positions. 
Replicating $M$ in a one dimensional lattice, we produce the infinite configuration space: $\widehat{ M}:=M\times \mathbb{Z}$.
Each of its cells is identified by an index 
$m \in \mathbb{Z}$: ${\widehat{ M}_m} := [0,1]  \times \{ m \}$. 
Every cell ${\widehat{ M}_{m}}$ contains two \emph{``slicers''}, 
$\{\ell_m\}\times\left\{m\right\}$ and $\{1-\ell_m\}\times\left\{m\right\}$, 
with $0 < \ell_m < 1/2$.
The slicers split each half of ${\widehat{M}_{m}}$ into two parts.
\citet{Salari} parameterised the value of $\ell_m$ by a positive number $\za$ as follows:
\begin{equation} \label{eq:SLICEGEN}
  \ell_m(\za) 
  = 
  \frac{1}{\left(\left|m\right|+2^{1/\za}\right)^{\za}} \, ,
  \qquad\text{with}\quad  m \in \mathbb{Z} \, , \quad \za > 0 \, .
\end{equation}
The SM, $S_\za : \widehat{M} \to \widehat{M}$, is then defined on the configuration space 
$\widehat{M}:=[0,1]\times\mathbb{Z}$ as follows:
\begin{equation}\label{eq:MAPPAGEN}
  S_\za(x,m)
  = 
  \left\{
    \begin{array}{rrl}
      (x,m-1) & \quad\text{for } & 0\leq x \le  \ell_{m} \text{ or } \frac{1}{2} < x \leq 1-\ell_{m},\\[2mm]
      (x,m+1) & \quad\text{for } & \ell_{m} < x \leq \frac{1}{2} \text{ or } 1-\ell_{m} < x \leq 1.
    \end{array}
  \right.
\end{equation}
The map is neither injective nor surjective.
It is nevertheless possible to define the inverse map 
when restricting to trajectories with initial conditions in cell $\widehat{M}_0$ \cite{Salari}.

The space $\widehat{M}$ can be endowed with a density of points that evolves under the action of $S_\za$. 
In particular, we consider the initial density 
$\hat{\mu} := \lambda\times\delta_{0}$ on $\widehat{M}$, 
where $\lambda$ is the Lebesgue measure on $M$ and $\delta_{0}$ is the Dirac measure on the integer $0$. 
Then, $S_\za$ can be interpreted as describing the transport of non-interacting particles in a one-dimensional space.%
\footnote{The ``particles'' are the points moved by $S_\za$. 
  Analogously to the particles of systems such as the Ehrenfest gas, 
  they do not interact with each other, since there is no coupling term connecting various 
  particles in their equations of motion.}

Let $\pi_{[0,1]}$ and $\pi_{\mathbb{Z}}$ be the projections of $\widehat{M}$ on its first and second factors, respectively.
Taking $x \in [0,1]$ and $m\in \Z$, we denote by $\hX=(x,m)$ a point in $\widehat{M}$, 
so that $\pi_{[0,1]} \hX = x$ and $\pi_{\mathbb{Z}} \hX = m$. 
Following \citet{Salari} we restrict our considerations to the initial distribution $\hat{\mu}$.
We view $\widehat{M}$ as subdivided in two halves that are invariant for the SM:
$\widehat{M}^+:=([1/2, 1]\times \{0\})\cup ([0,1] \times \Z^+)$ and 
$\widehat{M}^-:=([0, 1/2)\times \{0\})\cup ([0,1] \times \Z^-$).
The dynamics in the two intervals are the mirror images of each other. 
Indeed, since at $m=0$ the two slicers coincide with the single 
$\ell_0=1/2$, \cf~\Eq{SLICEGEN}, 
the points that lie initially in $[1/2, 1]$ never reach negative $m$, 
and those initially in $[0, 1/2)$ never reach positive $m$. 
Therefore, without loss of generality
we restrict the following analysis to the positive part of the chain, $\widehat{M}^+$. 
The sequence of integers $\pi_{\Z}(S^{j}(\hX)),\, j\in \N$, 
will be called the \emph{coarse-grained trajectory} of~$\hX$. The distance travelled by $\hX=(x,0)$ at time $n$
will be denoted  $\Delta \hX_{n}$.

A crucial aspect of the dynamics ${S}_\za$ is that its trajectories do not separate exponentially
in time. Indeed different trajectories in $\widehat{M}$ neither converge nor diverge from each 
other in time, except when (in a discrete set of points) they are separated by a slicer, and 
their distance jumps discontinuously.
\\%------------------------------------------------------}

%%%%  ------------------------------------------------------------------------------------------   %%%%
\subsection[Mean Maximum Displacements]{Mean Maximum Displacement and Maximum Square Displacement}

To illustrate some fundamental properties of the slicer dynamics, let us introduce the symbols
\begin{equation}\label{eq:deflp}
  \ell^{+}_{m}(\za)
  := 
  1 -\ell_{m}(\za) 
  = 
  1 - \frac{1}{(m+2^{1/\za})^{\za}} 
  \qquad\text{with}\quad
  m \in \N\cup\{0\} \, .
\end{equation}
They obey
\begin{equation}
  \frac 1 2 
  = \ell^{+}_{0}(\za) 
  < \ell^{+}_{1}(\za)
  < \dots  
  < \ell^{+}_{k}(\za) 
  < \ell^{+}_{k+1}(\za)
  < \dots 
  < 1,\quad 
  \mbox{and} \quad 
  \lim_{k\to \infty}  \ell^{+}_{k}(\za) 
  = 1 \, .
\end{equation}
Hence, there is a unique natural number $\bm = \bm_\za(x) > 0$ for any $x \in [1/2,1)$ 
such that
\begin{equation}\label{eq:LXL}
  \ell^{+}_{\bm-1}(\za)  
  < x 
  \le  \ell^{+}_{\bm}(\za) \, .
\end{equation}
In other words: 
\begin{equation}\label{eq:DEFFM}
  \bm_{\za}(x) 
  = 
  \min \{m\in \N  :   \ell^{+}_{m}(\za) \geq x  \} \quad \text{ for }  x \in [1/2,1) \, .
\end{equation}
Inspection of \Eq{MAPPAGEN} and the definition \eq{deflp} reveals that $\bm_\za(x)$ is
the maximum travelled distance for trajectories starting in the interval, \Eq{LXL}:
\vskip 5pt \noindent
\begin{lem}\label{lem:max-distance}{\bf \hskip -5pt :}
  Given $x \in [1/2, 1)$, let $\bm(x)$ be the integer that satisfies \Eq{LXL}.  Then,
  \begin{equation} 
    S_{\za}( x, \bm_\za(x) )
    = 
    (x, \bm_\za(x)-1),
    \qquad 
    S_{\za}( x, \bm_\za(x)-1 )
    =
    (x, \bm_\za(x)).
  \end{equation} 
\end{lem}
\vskip 5pt
\noindent
\proof This is a  straightforward consequence of \Eqs{MAPPAGEN} and \eq{LXL}. \hfill $\Box$
\vskip 5pt \noindent
This means that all trajectories become periodic with period $2$ after the  number $\bm_\za(x)$ of steps.
The description of the trajectory $\{ S_\za^j(\hX)\}_{j=0}^\infty$ with initial condition 
$\hX \in  \widehat{M}_{0} $ is completed by the following Proposition.
%% ----------------------------------------------------------------------------------------------------- %%
\vskip 5pt \noindent
\begin{prop}\label{prop:orbit}{\bf \hskip -5pt :}
For $x \in [1/2,1)$, let $\hX_{0} =(x,0) \in \widehat{M}_{0}$ and $\bm_\za(x)$ as defined by \Eq{DEFFM}.
Then the iterations of the trajectory starting at $\hX_{0}$ obey:
\begin{subequations} \label{eq:traject}
\begin{equation}
S_\za^k(x,0) = \left\{
  \begin{array}{llc@{\:\leq\:}l}
    (x, k)         & \quad\text{for } &  0          & k < \bm_{\za}(x) \, ,  \\[1mm]
    (x, \bt{k})    & \quad\text{for } & \bm_{\za}(x) & k \, ,
  \end{array}
\right .
\end{equation}
where 
\begin{equation}
\bt{k} = \left\{
  \begin{array}{lll}
    \bm_{\za}(x)    & \quad\text{for } & \left( k - \bm_{\za}(x) \right) \text{ is even} \, ,  \\[1mm]
    \bm_{\za}(x)-1  & \quad\text{for } & \left( k - \bm_{\za}(x) \right) \text{ is odd}  \, .
  \end{array}
\right .
\end{equation}
\end{subequations}
\end{prop}
\vskip 5pt \noindent
\proof 
This is a consequence of \Lemma{max-distance} and \Eq{MAPPAGEN}.
As long as $k < \bm_{\za}(x)$, the forthcoming iteration with $S_\za$ increases the cell index by one.
For $k \geq \bm_{\za}(x)$ the trajectory alternates between the cells $\bm_{\za}(x)$ and $\bm_{\za}(x)-1$.
\hfill$\Box$
%% ----------------------------------------------------------------------------------------------------- %%
\vskip 4pt \noindent
\begin{oss}
\Lemma{max-distance} and \Propo{orbit} imply that every trajectory starting at $\hX$ with $\pi_{[0,1]}(\hX) \in [1/2,1)$ 
is ballistic for a finite time, 
and then it gets localised eventually, turning periodic of period $2$.
\end{oss}
%% ----------------------------------------------------------------------------------------------------- %%
\begin{oss}
The trajectories starting at $\hX$ with $\pi_{[0,1]}(\hX)=1/2$ or $\pi_{[0,1]}(\hX)=1$ do not satisfy \Eq{LXL}.
Hence, they are forever ballistic, but they constitute a set of zero measure.
\end{oss}
\vskip 8pt
%% ----------------------------------------------------------------------------------------------------- %%

To investigate  the transport properties of the SM,  we observe that the function 
\begin{equation}
  \bm_\za(x):
  \left( 1/2, \,\,1 \right)
  \rightarrow
  \mathbb{N}
\end{equation}
is a step function with unitary jumps at the points $\ell^{+}_{m}(\za)$, such that
\begin{equation}\label{eq:steps}
  x  \in  (\ell^{+}_{k-1}(\za), \ell^{+}_{k}(\za)] 
  \mapsto   \bm_{\za}(x)
  =         k \, .
\end{equation}
Then, the following properties are satisfied:
\begin{enumerate}
\item $\bm_\za(x)$ is not decreasing: $x_{1}<x_{2}$ implies $\bm_{\za}(x_{1})\le \bm_{\za}(x_{2})$,
\item 
  $\bm_\za(x)$ is left continuous: 
  $\lim_{h \to 0^{-}}  \bm_{\za}(\ell^{+}_{k}(\za) + h) 
  = \bm_{\za}( \ell^{+}_{k}(\za) )
  = k$,
\item $\lim_{x\to 1/2^{+}} \bm_{\za}(x)=1$,\, 
  $\lim_{x\to 1^{-}} \bm_{\za}(x)=\infty$,
\item 
  $\za_{1} < \za_{2}$ implies $\bm_{\za_{1}}(x) \geq \bm_{\za_{2}}(x)$, since
  $\ell^{+}_j(\za_1) > \ell^{+}_{j}(\za_2)$ for $j > 0$.
\end{enumerate}
The points belonging to a strip
$(\ell^{+}_{k-1}(\za), \ell^{+}_{k}(\za)]  = \bm^{\,\,-1}_\za(k)$ 
share the same fate. 
Hence, the transport properties of the SM depend on the rate at which such strips shrink with growing~$k$.

Indeed, an ensemble of initial conditions $\widehat{E}_{0}$, \ie\ a set of points contained
in $(1/2, x_{0})\times \{0\}\subset \widehat{M}_{0}$ with $x_{0}<1$,
represents a coarse-grained version of the Dirac $\delta$ initial distribution, 
as commonly considered in diffusion theory.
This ensemble reaches localisation: the set $\{\pi_{\Z}(S_\za^j( \widehat{E}_{0})), j\in \N_0\}$ is bounded.
After all, the travelled distance does not exceed $\bm_{\za}(x_{0})$, which is finite. 
Consequently, non-trivial transport properties \emph{necessarily} require 
the initial ensemble $\widehat{E}_{0}$ to obey the condition:
\begin{subequations}
\begin{equation}\label{eq:SUPM}
\sup_{x \in \pi_{[0,1]}(\widehat{E}_{0})}  \bm_{\za}(x)=\infty,\, 
\end{equation}
or, equivalently, to accumulate at $x=1$:
\begin{equation}\label{eq:SUPE}
\sup(\pi_{[0,1]}(\widehat{E}_{0}))=1\,,
\end{equation}%
\end{subequations}%
however, note that condition \eqref{eq:SUPM} (or \eqref{eq:SUPE}) is not sufficient for non trivial behaviour, 
see Remark \ref{remark:primo}. Then in order to study transport,  we take an ensemble of \emph{uniformly 
distributed} initial
conditions $\widehat{E}_{0} \subset (1/2,1) \times \{ 0 \}$ \@{}---\@{}analogous to the setting in 
\citet{BCV10}\@{}---\@{}and 
characterise the transport properties of the SM by computing the corresponding ensemble averages.
Then, $\bm_{\za}(x)$ is the distance travelled by the point $\hX \in \widehat{E}_{0}$, with $\pi_{[0,1]}(\hX)=x$.
Consequently, the \emph{mean maximum displacement} and the \emph{mean maximum square displacement} are given by
\begin{equation}\label{eq:aveave}
  \langle \max_n \Delta \hX_n\rangle
  =
  \frac{1}{\lambda(\pi_{[0,1]}(\widehat{E}_{0}) )}  \hspace {-2 pt}
  \int\limits_{\pi_{[0,1]}(  \widehat{E}_{0} )}\!\!\! \hspace {-3 pt} \bm_{\za}(x) \, dx, 
  \; \mbox{and}\quad
  \langle \max_n \Delta \hX_n^{2}\rangle
  =
  \frac{1}{\lambda(\pi_{[0,1]}(\widehat{E}_{0}) )}   \hspace {-3 pt}
  \int\limits_{\pi_{[0,1]}(  \widehat{E}_{0} )} \!\!\! \hspace {-3 pt} \bm^2_{\za}(x) \, dx \, ,
\end{equation}
respectively, where
$\lambda(\pi_{[0,1]}(\widehat{E}_{0})) \leq 1/2$ is the Lebesgue measure of the projection of $\widehat{E}_{0}$
on $[0,1]$. Here and in the following, we denote by $\langle \cdot \rangle$ the ensemble average,
\ie\ the average with respect to the Lebesgue measure normalized on $\pi_{[0,1]}(\widehat{E}_{0})$,
and we assume that $\widehat{E}_{0}= (1/2, 1)\times \{0\}$.

These averages do not depend on time.
However, they indicate what can be expected for the time evolution of the average travelled distance and mean-square distance.
To understand this point, we observe that in each interval 
$(\ell^{+}_{k-1}(\za), \ell^{+}_{k}(\za)]$, $k \in \N$ 
the function $\bm_{\za}(x)$ takes the constant value $k$.
We denote the length of these intervals by:
\begin{subequations}
\begin{equation}\label{eq:def_delta}
  \Delta_{k}(\za) := \ell^{+}_{k}(\za)-\ell^{+}_{k-1}(\za) \, .
\end{equation}
By construction their length adds up to $1/2$,
\begin{equation}\label{eq:unmezzo}
  \sum_{k=1}^{\infty} \Delta_{k}(\za) 
  =  
  \frac 1 2, 
\end{equation}
and to leading order in $k$, we have:
\begin{equation}\label{eq:lenDeltaAlpha}
   \Delta_{k}(\za)
   = 
   \frac{\za}{k^{\za+1}}
   \left ( 1
           - \frac{\tilde{c}(\za) }{k} 
           + O(k^{-2})
   \right)
   \qquad\mbox{with }\quad 
   \tilde{c}(\za) = (1+\za) \, \left( 2^{1/\za} - \frac{1}{2} \right) \, .
\end{equation}
\end{subequations}
Then, recalling that $\widehat{E}_{0}= (1/2, 1)\times \{0\}$, one finds
\begin{subequations}
\begin{equation}\label{eq:series}
  \langle \Delta \hX\rangle
  =
  2 \int\limits_{1/2}^{1}\bm_{\za}(x) \, dx
  = 2 \sum_{k=1}^{\infty} k \, \Delta_{k}(\za)
  = 2 \sum_{k=1}^{\infty}  \frac{\za}{k^{\za}} \, \left( 1 + O(k^{-1}) \right),
\end{equation}
where \Eqs{steps} and \eq{lenDeltaAlpha} have been used. 
Thus, $\langle \Delta \hX\rangle$ converges for $\za >1$, and it diverges otherwise. 
Analogously, the mean maximum square displacement is
\begin{equation}\label{eq:amsd}
  \langle \Delta \hX^{2}\rangle
  = 
  2 \int\limits_{1/2}^{1}\bm_{\za}^{2}(x) \, dx
  =
  2 \sum_{k=1}^{\infty} k^{2} \Delta_{k}(\za)
  =  
  2 \sum_{k=1}^{\infty}  \frac{\za}{k^{\za-1}} \, \left( 1 + O(k^{-1}) \right) \, .
\end{equation} 
\end{subequations}
For $\za>2$ the square displacement, $\langle \Delta \hX^{2}\rangle$, is finite. 
This corresponds to the localisation phenomenon described in Remark~6 of \citet{Salari}. 
It arises from the fact that $\ell_k^+(\za)$ tends to $1$ faster, 
and transport of the SM is slower, for larger $\za$.
On the other hand, for $ 0<\za<2$ the mean maximum square displacement diverges, 
and it is of interest to explore the rate at which this divergence takes place, 
\ie\ to determine the \emph{transport exponent}~$\zg$.

%%%%  ------------------------------------------------------------------------------------------   %%%%
\subsection{Time Evolution of the Displacement Moments}
Each particle moves by exactly one step in each time step.
Hence, trajectories reach at most site $n$ in $n$ time steps, 
and the distance $\Delta \hX_{n}$ travelled by $\hX=(x,0)$ at time $n$ is given by
\begin{equation}
  \min \{\bt{n},n \} \, , 
\end{equation}
\cf~\Eq{traject}.
Moreover, for even and odd times $n$ the displacement $\Delta \hX_{n}$ also takes even and odd values, respectively.
The corresponding (time-dependent) mean-square displacement can be written as
\begin{subequations}
\begin{eqnarray}
  \langle \Delta \hX^{2}_{n}\rangle 
  &=& 
  2 \int\limits_{1/2}^{1}  \min \{\bt{n}, n \}^{2} \, dx 
\nonumber\\[2mm]
  &=& \left\{ \begin{array}{lll}
                \displaystyle
                2 \sum_{i=1}^{(n/2)-1} (2i)^{2} \left( \Delta_{2i}(\za) + \Delta_{2i+1}(\za) \right)
                + 2 n^{2} \sum_{k=n}^{\infty} \Delta_{k}(\za) 
                & \text{for}
                & $n$ \text{ even} \, ,
                \\[3mm]
                \displaystyle
                2 \sum_{i=1}^{(n-1)/2} (2i-1)^{2} \left( \Delta_{2i-1}(\za) + \Delta_{2i}(\za) \right)
                + 2 n^{2} \sum_{k=n}^{\infty} \Delta_{k}(\za) 
                & \text{for}
                & $n$ \text{ odd} \, .
              \end{array} \right .
\end{eqnarray}
The sums involving terms $k \geq n$ collect the particles that make $n$ steps to the right and never turned back. 
\citet{Salari} denoted this as the travelling area.
It is the same in both cases.
The other sum accounts for particles that turn back at least once.
Consequently, the particles get localised within $n$ time steps.
This represents the term called sub-travelling area in \citet{Salari}.
To leading order this contribution to the mean-square displacement takes the same for odd and even $n$.
Hence, we write:
\begin{eqnarray}
  \langle \Delta \hX^{2}_{n}\rangle 
  = 
  2 \int\limits_{1/2}^{1}  \min \{\bt{n}, n \}^{2} \, dx 
  = 2 \sum_{k=1}^{n-1} k^{2} \Delta_{k}(\za) \: \left( 1 + O(k^{-1}) \right)
  + 2 n^{2} \sum_{k=n}^{\infty} \Delta_{k}(\za) \, .
\end{eqnarray}%
\label{eq:intMSD}%
\end{subequations}%
The asymptotic behaviour of the first sum is:%
\footnote{By $f_1(n)\sim f_2(n)$ we mean $f_1(n)/f_2(n)\to 1$ as $n\to \infty$.}
\begin{subequations}
\begin{equation}\label{eq:sumDeltaKsquareScaling}
  2 \, \sum_{k=1}^{n-1} k^{2} \Delta_{k}(\za) 
  =
  2 \, \sum_{k=1}^{n-1} \frac{\za}{k^{\za-1}} \, \left( 1 + O(k^{-1}) \right)
  \sim 
  \left\{
  \begin{array}{lll}
    \frac{2 \, \za}{2-\za} n^{2-\za}
    & \quad\text{for }
    & 0 < \za < 2 \, ,
    \\[1mm]
    4 \, \ln n 
    & \quad\text{for }
    & \za = 2 \, ,
    \\[1mm]
    \text{const}
    & \quad\text{for }
    & \za > 2 \, .
  \end{array}
  \right.
\end{equation}
The form of this scaling can be guessed by interpreting the sum as a Riemann-sum approximation of the integral 
$\int_1^n x^{1-\za} \, dx$. 
A formal derivation is given in Appendix~\ref{app:SumScaling}. 
The second sum can be evaluated based on the definition of $\Delta_{k}(\za)$,
\begin{equation}\label{eq:sumDeltaScaling}
  2 \, n^2 \sum_{k=n}^{\infty} \Delta_{k}(\za)
   =
  2 \, n^2 \, \ell_{n-1}(\za) 
  =
  2 \, n^2 \, n^{-\za} \left( 1 - \frac{\za \, 2^{1/\za}}{n} + O(n^{-2}) \right)
  \sim 
  2 \, n^{2-\za}
  \, .
\end{equation}%
\end{subequations}%
%% ----------------------------------------------------------------------------------------------------- %%
\begin{oss}
According to \Eqs{sumDeltaKsquareScaling} and \eq{sumDeltaScaling}
the travelling and the sub-travelling areas have the same asymptotic scaling. 
\end{oss}
%% ----------------------------------------------------------------------------------------------------- %%
Altogether, we find that the mean-square displacement scales like
\begin{equation}\label{eq:MeanSquareDisplacement}
  \langle \Delta \hX^{2}_{n}\rangle  
  \sim 
  \left\{
  \begin{array}{lll}
    \frac{4}{2-\za} \: n^{2-\za}
    & \quad\text{for }
    & 0 < \za < 2 \, ,
    \\[1mm]
    4 \, \ln n
    & \quad\text{for }
    & \za = 2 \, ,
    \\[1mm]
    \text{const}
    & \quad\text{for }
    & \za > 2 \, .
  \end{array}
  \right .
\end{equation} 
The computation of other moments 
$\langle |\Delta \hX_{n}|^{p}\rangle$, with $p > \za$,
can be obtained in the same way, based on the same integral representation:
\begin{eqnarray}
  \langle |\Delta \hX_{n}|^{p}\rangle 
  &=& 
  2 \int\limits_{1/2}^{1}  \min\{\bt{n}, n \}^{p} \, dx 
  \; \sim \;
    2  \sum_{k=1}^{n-1} k^{p} \Delta_{k}(\za) \left( 1 + O(k^{-1}) \right)
  + 2 n^{p} \sum_{k=n}^{\infty} \Delta_{k}(\za) 
  \nonumber\\[2mm]\label{eq:DeltaMoments}
  & \sim & 
  \left\{
  \begin{array}{lll}
   \frac{2\, p}{p-\za} \; n^{p-\za}
    & \quad\text{for }
    & 0 < \za < p \, ,
    \\[1mm]
    2\, p \, \ln n
    & \quad\text{for }
    & \za = p \, ,
    \\[1mm]
    \text{const}
    & \quad\text{for }
    & \za > p \, ,
  \end{array}
  \right .
\end{eqnarray}%
because
\begin{equation}\label{eq:k-to-p}
  \sum_{k=1}^{n-1} k^{p} \Delta_{k}(\za) 
  =
  2 \, \sum_{k=1}^{n-1} \frac{\za}{k^{\za-p+1}} \, \left( 1 + O(k^{-1}) \right)
  \sim 
  \left\{
  \begin{array}{lll}
    \frac{2 \, \za}{p-\za} \; n^{p-\za}
    & \quad\text{for }
    & 0 < \za < p \, ,
    \\[1mm]
    2\, p \, \ln n
    & \quad\text{for }
    & \za = p \, ,
    \\[1mm]
    \text{const}
    & \quad\text{for }
    & \za > p \, .
  \end{array}
  \right .
\end{equation}
We hence reproduced central results of \citet{Salari} in a formalism 
that is suitable to compute the position-position auto-correlation function.
These findings are summarised by the following theorem.
%% ----------------------------------------------------------------------------------------------------- %%

\begin{thm}\hskip -4pt{\rm \bf :}\label{thm:teorema1}
  Given $0 \le \za <2$, the transport exponent of the Slicer Dynamics
  with uniformly distributed initial condition in $\widehat{M}_0$ takes the value $\gamma=2-\za$,
  and the behaviour is
\begin{enumerate}
\item ballistic if $\za=0$,
\item super-diffusive if $0<\za <1$,
\item diffusive  if $\za=1$,
\item sub-diffusive if $1< \za <2$,
\item logarithmically growing for the mean-square displacement, $\langle \Delta \hX^{2}_{n}\rangle \sim \ln n$, if $\za = 2$.
\end{enumerate}
For $\alpha > 2$  the dynamics has
\begin{enumerate}
\setcounter{enumi}{5}
\item 
  bounded  mean-square displacement  $\langle \Delta \hX^{2}_{n}\rangle$.
\end{enumerate}
Furthermore, for $p>\za$ the moments satisfy $\langle |\Delta \hX_{n}|^{p}\rangle  \sim n^{p-\za}$.
\end{thm}

%% ----------------------------------------------------------------------------------------------------- %%
\vskip 4pt \noindent
\begin{oss}\hskip -4pt{\rm \bf :}\label{remark:primo}
  The parameter dependence of the transport exponent $\gamma$ 
  depends on the initial distribution. 
\end{oss}
%% ----------------------------------------------------------------------------------------------------- %%
For instance, suppose that the $x$-component of the initial conditions has got density $\rho$ with respect 
to the uniform measure $dx$ in the interval $(1/2, 1)$.
Then, in place of \Eq{intMSD} we have 
\begin{equation}\label{eq:intMSDrho}
  \langle \Delta \hX^{2}_{n}\rangle_\rho 
  = 2 \int\limits_{1/2}^{1}  \min \{\bt{n}, n \}^{2} \; \rho(x) \, dx \, .
\end{equation}
If the support of $\rho$  does not contain a (left) neighbourhood of $1$, 
then $\langle \Delta \hX^{2}_{n}\rangle \to \text{const}$ as $n\to \infty$ even for $\za < 1$. 
Hence, different asymptotic behaviours arise from different distributions supported in a neighbourhood of $1$. 
For instance, take $\rho(x) = O((1-x)^r)$ as $x\to 1^-$ with $r>-1$. 
Then, different values of $r$ produce different kinds of diffusion, even at fixed $\za$. 
Specifically, for the ensemble
\begin{equation}\label{eq:rho}
  \rho(x) = \left \{ 
    \begin{array}{lll}
       r\, 2^{r-1} \, \left( \frac{1}{2} - x \right)^{r-1} 
      & \quad\text{for }
      & 0 \leq x \leq \frac{1}{2} \, ,
      \\[2mm]
      r\, 2^{r-1} \, \left( 1-x \right)^{r-1} 
      & \quad\text{for }
      & \frac{1}{2} < x \leq 1 \, ,
    \end{array}
  \right .
\end{equation}
one  finds that
\begin{equation}\label{eq:MeanSquareDisplacementRho}
  \langle \Delta \hX^{2}_{n}\rangle_\rho 
  \sim 
  \left\{
  \begin{array}{lll}
    \frac{2^{r+1}}{2-r \za} \: n^{2-r \za}
    & \quad\text{for }
    & 0 < r\,\za < 2 \, ,
    \\[1mm]
    2^{r+1}\, \ln n
    & \quad\text{for }
    & r\, \za = 2 \, ,
    \\[1mm]
    \text{const}
    & \quad\text{for }
    & r\, \za > 2 \, .
  \end{array}
  \right .
\end{equation} 
The transport exponent $\zg = 2-r\,\za$ depends continuously on $r$, 
and only for $r=1$ (the case of the uniform distribution treated so far) 
does \Eq{MeanSquareDisplacementRho} reduce to \Eq{MeanSquareDisplacement}.
Different initial distributions lead the SM to different transport properties, as 
already observed in other frameworks, such as those of L\'evy walks \cite{BCV10}. 
However, the dynamical mechanisms underlying
this finding are drastically different.

\vskip 4pt \noindent
\begin{oss}\hskip -4pt{\rm \bf :}\label{remark:memory}
  The anomalous behaviour of the SM with $\za \ne 1$
  coincides with the persistence of memory. 
  This is in accordance with the common observation in thermodynamic systems that
  slow decay of correlation leads to anomalous transport,  
  while rapid decay of correlations leads to normal diffusion.
  However, for $\za=1$ the statistics of the SM appear still like normal diffusion,
  although memory persists in this case,
  just as it does for $\za \ne 1$.
  Hence, the SM illustrates that a statistical coincidence in certain phenomena should not
  be taken as a thermodynamic phenomenon
  when no thermodynamics is present \citep[see][for another example]{Cohen1}. 
\end{oss}
%% ----------------------------------------------------------------------------------------------------- %%
In the case of the SM, the term {\em transport} must then be used with a grain of salt.

In the following, we explore whether correlations might help to distinguish the SM from the LLg. 
To this end we analytically compute various 
scaling limits of the position-position auto-correlations generated by the SM, 
and compare them with numerically computed correlations of the LLg with the same exponent $\zg$.

%%%%  ------------------------------------------------------------------------------------------   %%%%
\subsection[Correlations in the Slicer Dynamics]{Position-Position Correlations in the Slicer Dynamics}
\label{sec:corrslicer}
Let us introduce the position-position auto-correlation function as:
\begin{subequations}
\begin{eqnarray}\label{eq:autocorrelation}
  \phi(n,m) 
  &:=& \langle\pi_\mathbb{Z}(S^n(\hX)) \; \pi_\mathbb{Z}(S^m(\hX)) \rangle 
  := \langle
       \Delta \hX_{n} \,\Delta \hX_{m} \rangle 
  \\[2mm]
  &=& 2 \int\limits_{1/2}^1 
      \min\{\bt{m}, m\}\, \min\{\bt{n}, n\}
      \: dx  \qquad \text{with } m \leq n \, .
\end{eqnarray}
\end{subequations}
The integration interval
  $\mathrm{I} := (1/2, 1]$ 
can be subdivided in three parts, 
  $\mathrm{I} = E^<_m\cup E_{m,n}\cup E^>_n$,
defined by
\begin{equation}\label{eq:setsE}
\begin{array}{rll}
E^<_m  &= \{x \in \mathrm{I}\, :\, \bt{m} \leq m \}
      & \Rightarrow\;  \min\{\bt{m}, m\}\, \min\{\bt{n}, n\} = \bt{m} \, \bt{n} \, ,
\\[2mm]
E_{m,n} &= \{x \in \mathrm{I}\, :\, m < \bt{n} \leq n\}
       & \Rightarrow\;  \min\{\bt{m}, m\}\, \min\{\bt{n}, n\} = m \; \bt{n} \, ,
\\[2mm]
E^>_n  &= \{x\in \mathrm{I}\, :\, n < \bt{n} \} 
       & \Rightarrow\;  \min\{\bt{m}, m\}\, \min\{\bt{n}, n\} = m \, n \, .
\end{array}
\end{equation}
Then, rewriting the resulting integrals in terms of sums over the intervals 
where $\bm_\za(x)$ takes the constant value $k$ (\cf~\Eq{intMSD}), one has:
\begin{subequations}\label{eq:slicerCorrelation}
\begin{eqnarray}\label{eq:slicereq8a}
\hskip -20pt
\phi(n,m) 
&=& 2 \int\limits_{E^<_m}     \bt{n} \: \bt{m} \, dx
  + 2\, m \int\limits_{E_{m,n}} \bt{n}   \, dx 
  + 2\, m\,n \int\limits_{E^>_n}            \, dx
\label{eq:intgrl corr.expr.}
\\[2mm] 
\label{eq:corr.expr.}
& \sim &
            2 \sum\limits_{k=1}^{m} k^2 \, \Delta_k(\za) 
          + 2\, m \sum\limits_{k=m+1}^{n} k \, \Delta_k(\za) 
          + 2 \, m\,n \sum\limits_{k=n+1}^\infty \Delta_k(\za) \, ,
\quad m \leq n \, .
\end{eqnarray}
\end{subequations}
The first and the third sum have been evaluated in \Eqs{sumDeltaKsquareScaling} and \eq{sumDeltaScaling}, respectively.
The asymptotic behaviour of the second term depends on the value of $\za$ and on the relation between $m$ and $n$.
In the following, we discuss the following examples:
\begin{itemize}
\item[1. ]    $n\to \infty$ with $m$ fixed,
\item[2. ]   $n, m \to \infty$ with fixed $h = n-m$,
\item[3. ]  $n, m \to \infty$ with $n = m+\ell\, m^q$, where $\ell, q$ are positive constants.
\end{itemize}

%%%%  ------------------------------------------------------------------------------------------
\subsubsection{Scaling of $\phi(n,m)$ for $n \to \infty$ with $m$ fixed}
\label{sec:sub31} 

In order to evaluate the second sum in \Eq{corr.expr.} we observe that
\begin{eqnarray}\label{eq:2ndSumDecomposition}
  2\, m \sum\limits_{k=m+1}^{n} k \, \Delta_k(\za) 
  & = &
    2\, m \sum\limits_{k=0}^{n} k \, \Delta_k(\za) 
  - 2\, m \sum\limits_{k=0}^{m} k \, \Delta_k(\za) \, .
\end{eqnarray}
For fixed $m$, the latter sum takes a constant value, and for $n \to \infty$ the former sum scales as
(\cf~\Eq{k-to-p} or the formal derivation provided in  Appendix~\ref{app:SumScaling})
\begin{eqnarray}\label{eq:2m.sum.k.Delta-fix.m}
  2\, m \sum\limits_{k=0}^{n} k \, \Delta_k(\za) 
  \sim 
  \left\{
  \begin{array}{lll}
    \frac{2 \, \za\, m}{1-\za} \: n^{1-\za}
    & \quad\text{for }
    & 0 < \za < 1 \, ,
    \\[2mm]
    2\, m \, \ln n
    & \quad\text{for }
    & \za = 1 \, ,
    \\[1mm]
    \text{const}
    & \quad\text{for }
    & \za > 1 \, .
  \end{array}
  \right . 
\end{eqnarray}
The leading-order scaling of the three sums in \Eq{corr.expr.} is summarised in the following lemma:
\vskip 5pt \noindent
\begin{lem}\hskip -4pt{\rm \bf :}\label{thm:fixed-m-scaling}
  For $n \to \infty$ with fixed $m$ the auto-correlation function, $\phi(n,m)$, 
  defined in \Eq{autocorrelation}, asymptotically scales as:
\begin{equation}\label{eq:fixed-m-scaling}
  \phi(n,m) 
  \sim 
  \left\{
  \begin{array}{lll}
    \frac{2\, m}{1-\za} \: n^{1-\za}
    & \quad\text{for }
    & 0 < \za < 1 \, ,
    \\[2mm]
    2\, m \, \ln n
    & \quad\text{for }
    & \za = 1 \, ,
    \\[2mm]
    \text{const}
    & \quad\text{for }
    & \za > 1 \, .
  \end{array}
  \right . 
\end{equation}
\end{lem}
\vskip 5pt
\noindent
\proof 
The first sum in \eqref{eq:corr.expr.} has a finite number of terms that all take finite positive values.
Hence, it adds to a finite positive number.
For $0 < \za < 1$ the leading-order contributions of the second and the third sum have the same scaling, 
$n^{1 - \za}$, which diverges for $n \to \infty$. From \Eqs{2m.sum.k.Delta-fix.m} and \eq{sumDeltaScaling} we have
\[
\phi(n,m) 
\sim 2\, m \; \left( \frac{\za}{1-\za} + 1 \right) \: n^{1-\za}
= \frac{2\, m}{1-\za} \: n^{1-\za}
\qquad \text{for } 0 < \za < 1 \, .
\]
For $\za = 1$ the exponent $1-\za = 0$ such that the third term also takes a finite value.
In that case the leading-order scaling is provided by the second sum, \Eq{2m.sum.k.Delta-fix.m}.

Finally, for $\za > 1$ all sums contributing to \Eq{autocorrelation} take constant values.
\hfill $\Box$
\vskip 8pt \noindent

\begin{figure}
{\hfill
\subfloat[\label{fig:fix_m_theory} 
Correlations for fixed time $m$.]{\includegraphics[width=0.47\textwidth]{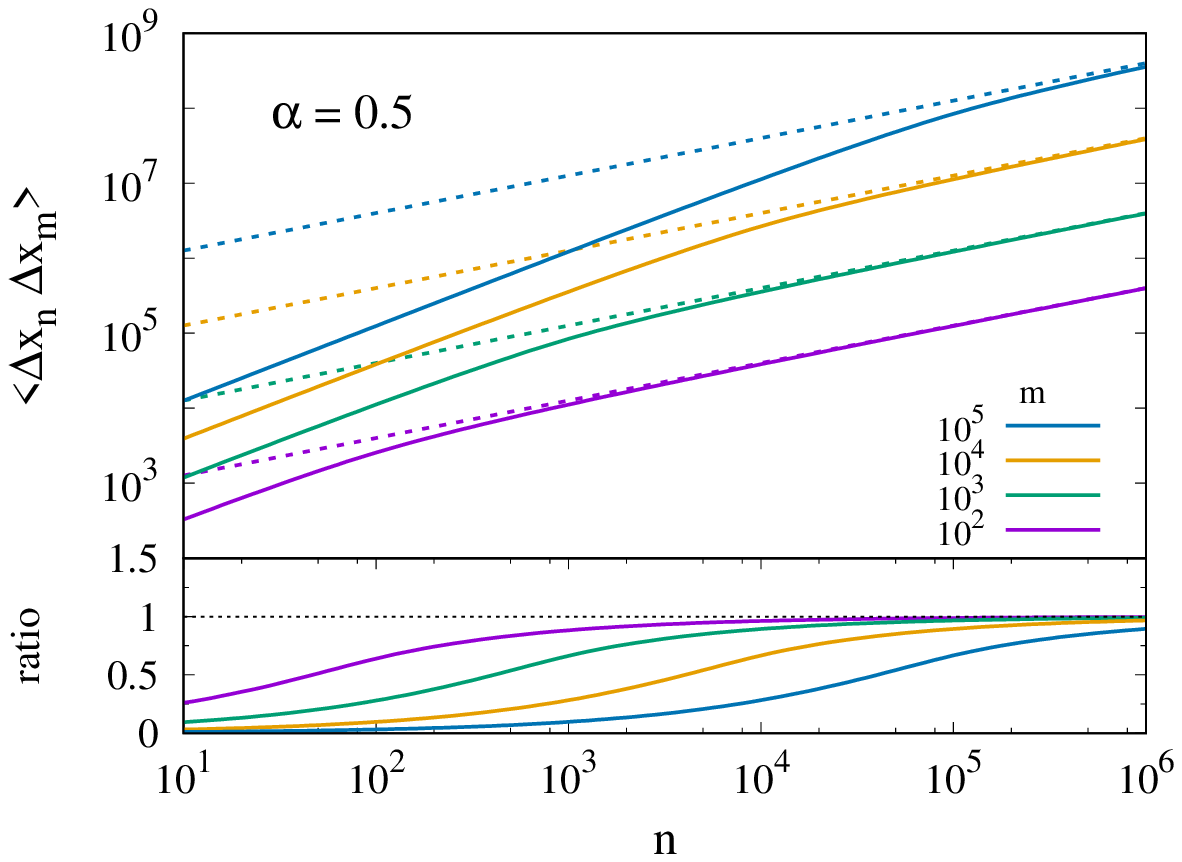}}
\hfill
\subfloat[\label{fig:fix_h_theory} 
Correlations for fixed time lag $h=n-m$.]{\includegraphics[width=0.47\textwidth]{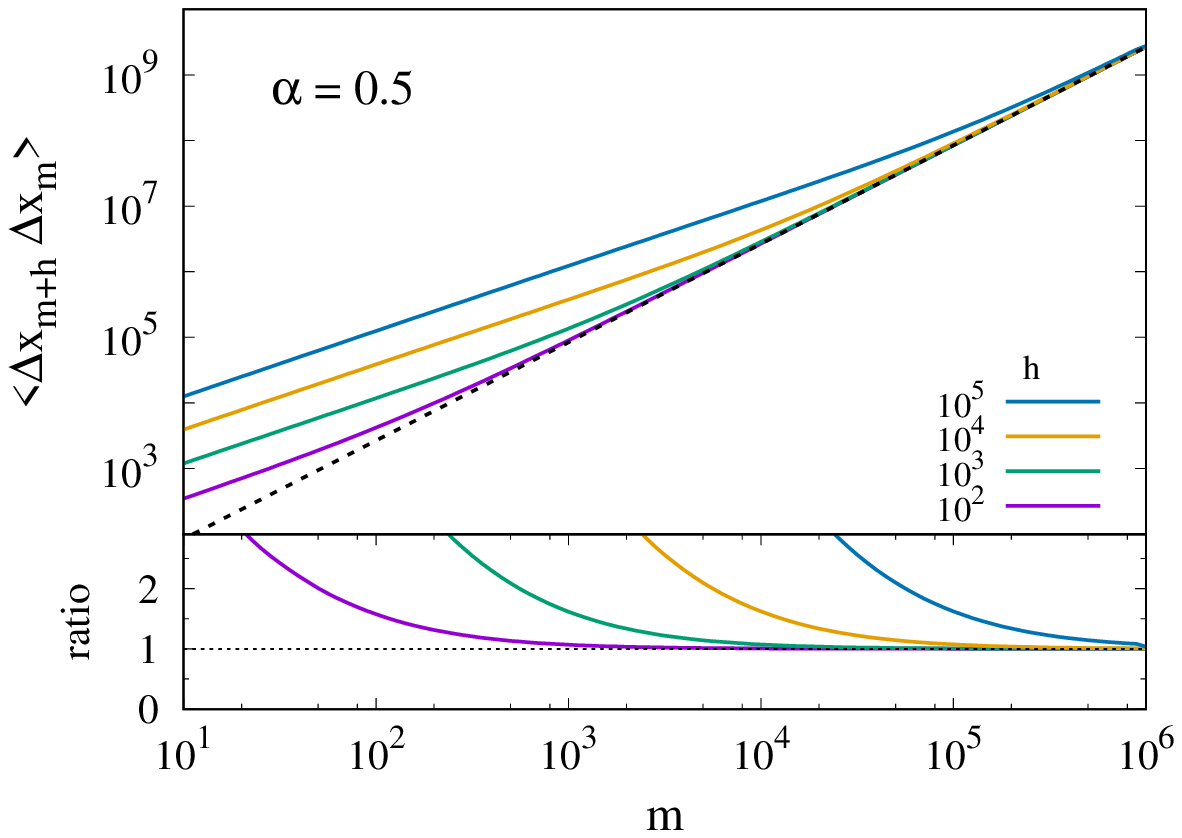}}
\hfill}
\caption[]{Comparison of the auto-correlation function, $\phi(n,m)$, 
and expressions for its asymptotic scaling, for $\za=1/2$ in two different cases. 
The functions $\phi(n,m)$ obtained from the sums of \Eq{corr.expr.} are plotted as solid lines.
The respective asymptotic expressions are indicated by dashed lines.
(a) The limit of large $n$ for a for fixed value of $m$.
The values of $m$ are provided in the figure legend in the order of the lines from top to bottom.
The asymptotic scaling is provided by \Eq{fixed-m-scaling}.
(b) The large-$m$ limit for a fixed time lag $h=n-m$.
The values of $h$ are provided in the figure legend in the order of the lines from top to bottom.
The asymptotic scaling is provided by \Eq{fixed-h-scaling}.
In the lower panels we show the ratio of the auto-correlation function and the respective asymptotic expressions.
}
\end{figure}

The dashed lines in Figure~\ref{fig:fix_m_theory} show the asymptotic behaviour, \Eq{fixed-m-scaling}, 
for $\za=1/2$ and different fixed values of $m$. 
They provide an excellent description of the asymptotic behaviour of
the numerical evaluation of the definition, \Eq{autocorrelation} (solid lines).
The lower panel of the figure demonstrates
that the ratio of the correlation function and the prediction of its asymptotic behaviour
approaches one for a vast range of different values of $m$.

%%  ------------------------------------------------------------------------------------------
%%  ------------------------------------------------------------------------------------------

%%%%  ------------------------------------------------------------------------------------------
\subsubsection{Scaling of $\phi(m+h, m)$ for  $m \to \infty$ with  $h > 0$ fixed}
\label{sec:sub32}

In this case, the second sum in \Eq{corr.expr.}  involves a finite number of positive terms.
The sum can be bounded from above by
\begin{align*}
  & \hskip -15pt 
  2 \, m \sum_{k=m+1}^{m+h} k\, \Delta_{k}(\za) 
  =  
  2 \, \za\, m \sum_{k=m+1}^{m+h} k^{-\za}\, 
        \left( 1 - \frac{\tilde{c}(\za) }{k} + O(k^{-2}) \right)
  \\[2mm]
  & \hskip 10pt 
  < 
  2 \, \za\, h \, m^{1-\za} \, \left( 1 - \frac{\tilde{c}(\za) }{m+h} + O(m^{-2}) \right)
   < 
        \left\{
        \begin{array}{lll}
          2 \, \za\, h \, (m+h)^{1-\za} 
          & \quad\text{for }
          & 0 < \za < 1 \, ,
          \\[2mm]
          2 \, \za\, h 
          & \quad\text{for }
          & 1 \leq \za \, .
        \end{array}
        \right .
\end{align*}
and from below by
\begin{align*}
  2 \, m \sum_{k=m+1}^{m+h} k\, \Delta_{k}(\za)
   > 
  2 \, \za\, h \, m \, (m+h)^{-\za} \, \left( 1 + O(m^{-1}) \right)
   > 
        \left\{
        \begin{array}{lll}
          2 \, \za\, h \, m^{1-\za} 
          & \quad\text{for }
          & 0 < \za < 1 \, ,
          \\[2mm]
          0 
          & \quad\text{for }
          & 1 \leq \za \, .
        \end{array}
        \right .
\end{align*}
Noting that constant $h$ implies $(m+h)^{1-\za} = m^{1-\za} \, ( 1+ h/m )^{1-\za}  \sim m^{1-\za}$
we find that the second sum scales as
\begin{equation}\label{eq:2m.sum.k.Delta-fix.h}
  2 \, m \sum_{k=m+1}^{m+h} k\, \Delta_{k}(\za)
  \sim
  \left\{
    \begin{array}{lll}
          2 \, \za\, h \, m^{1-\za} 
          & \quad\text{for }
          & 0 < \za < 1 \, ,
          \\[2mm]
          O(1)
          & \quad\text{for }
          & 1 \leq \za \, .
    \end{array}
  \right .
\end{equation}
Hence, the leading-order scaling of the auto-correlation function takes the form:
\vskip 5pt \noindent
\begin{lem}\hskip -4pt{\rm \bf :}\label{thm:fixed-h-scaling}
  For $m \to \infty$ with fixed $n-m = h = \text{const}$ the auto-correlation function, 
  $\phi(m+h, m)$, asymptotically scales as:
\begin{equation}\label{eq:fixed-h-scaling}
  \phi(m+h, m) 
  \sim 
  \left\{
  \begin{array}{lll}
    \frac{4}{2-\za} \: m^{2-\za}
    & \quad\text{for }
    & 0 < \za < 2 \, ,
    \\[2mm]
    4\, \ln( m )
    & \quad\text{for }
    & \za = 2 \, ,
    \\[2mm]
    \text{const}
    & \quad\text{for }
    & \za > 2 \, .
  \end{array}
  \right . 
\end{equation}
\end{lem}
\vskip 5pt
\noindent
\proof 
For $0 < \za < 2$ the leading-order contributions of the first and third term in \Eq{corr.expr.} 
have the same scaling, $m^{2-\za}$.
These terms dominate the scaling of the second sum, \Eq{2m.sum.k.Delta-fix.h}.
In this range one hence recovers the scaling of the mean-square displacement in \Eq{MeanSquareDisplacement}.

For  $\za = 2$ the second and third terms in \Eq{corr.expr.} take constant values, 
while the first one diverges logarithmically according to \Eq{sumDeltaKsquareScaling}. 

Finally, for $\za > 2$ all sums contributing to \Eq{corr.expr.} take constant values.
\hfill $\Box$
\vskip 8pt \noindent

The dashed lines in Figure~\ref{fig:fix_h_theory} show the asymptotic behaviour, \Eq{fixed-h-scaling}, for $\za=1/2$, and different time lags $h=n-m$.
They provide an excellent description of the asymptotics of the numerical evaluation of the definition, \Eq{autocorrelation} (solid lines).
The lower panel of the figure demonstrates that the ratio of the auto-correlation function 
and the prediction of its asymptotic behaviour approaches $1$ for a vast range of values of $h$.

The correlation function $\phi(m+h, m)$ looks like correlation functions addressing the time-translation invariance 
of the position-position auto-correlation function $\phi(t_1, t_2)$ for fixed time increments $t_2-t_1 = h$. 
However, this impression is misleading:
here, we consider an ensemble where all members start close to the origin.
Therefore, the behaviour of the correlations for large $m$ characterizes the decay of features of the initial ensemble
rather than referring to translations in time.
In order to clearly make this point we consider the large $m$ scaling of $\phi(m+\ell m^{q}, m)$.
For $q<1$ the difference of the two times will become negligible as compared to the mean.
For $q>1$ the difference between $t_1$ and $t_2$ grows.

%%%%  ------------------------------------------------------------------------------------------
\subsubsection{Scaling of $\phi(m+\ell m^{q}, m)$ for $m\to\infty$ with $\ell > 0$}
\label{sec:sub33}

For $q<1$, $q=1$, and $q>1$ the auto-correlation function shows different scalings.

\paragraph{Scaling for $q<1$. \ }

In this case bounds for the second sum in \Eq{corr.expr.} 
can be provided by a calculation fully 
analogous to the derivation of \Eq{2m.sum.k.Delta-fix.h}.
This provides the scaling
\begin{equation}
  2 \, m \sum_{k=m+1}^{m+h} k\, \Delta_{k}(\za)
  \sim
  \left\{
    \begin{array}{lll}
          2 \, \za\, \ell\, m^{1-q-\za} 
          & \quad\text{for }
          & 0 < \za < 1 \, ,
          \\[2mm]
          O(1)
          & \quad\text{for }
          & 1 \leq \za \, .
    \end{array}
  \right .
\nonumber
\end{equation}
This scaling is always sub-dominant with respect to those of the other two sums in \Eq{corr.expr.}.
As far as the asymptotic scaling is concerned we have the same situation as for fixed $n-m = h$,
and the auto-correlation function has the same scaling in these two limits.
\vskip 5pt \noindent
\begin{lem}\hskip -4pt{\rm \bf :}\label{thm:q.leq.1-scaling}
For  $q<1$, $\ell > 0$, and $m \to \infty$ the auto-correlation function, 
$\phi(m + \ell m^q, m)$ follows the same asymptotic scaling, \Eq{fixed-h-scaling},
as for the case where the time difference between the arguments is constant,
\begin{equation}
  \phi(m+\ell\,m^q, m) \sim \phi(m+h, m) 
  \qquad\text{for}\quad \ell, h > 0 
  \quad\text{and}\quad q < 1 \, .
\end{equation}
\end{lem}
\vskip 5pt \noindent

\paragraph{Scaling for $q=1$. \ }
In this case we have $n = m+\ell\,m^q = (1+\ell) m$, \ie\ $n$ is proportional to $m$.
In order to find the scaling for large $m$, we start from \Eq{2ndSumDecomposition}.
For $\za \neq 1$ the two sums on the right-hand side scale like a power law with exponent $1-\za$
and a constant offset that is relevant when $\za > 1$. 
The constant drops out when taking the difference, so that we obtain
\begin{subequations}
\begin{eqnarray}\label{eq:2m.sum.k.Delta-q}
  2\, m \sum\limits_{k=m+1}^{n} k \, \Delta_k(\za) 
  & \sim &
  \frac{2\, m \, \za}{1-\za} \:  \left( n^{1-\za} - m^{1-\za} \right)
  \\[2mm]
  \label{eq:2m.sum.k.Delta-q.eq.1A}
  &=&
  \frac{2\, \za}{1-\za} \: \left( (\ell + 1)^{1-\za} - 1 \right) \: m^{2-\za} 
  \qquad\text{for}\quad
  \za \neq 1 \, .
\end{eqnarray}
Moreover, for $\za = 1$ the sum diverges logarithmically:
\begin{eqnarray}\label{eq:2m.sum.k.Delta-q.eq.1B}
  2\, m \sum\limits_{k=m+1}^{n} k \, \Delta_k(\za) 
  \sim
  2\, m \, \za \:  \ln\frac{n}{m}
  =
  2\, m\:  \ln(1+\ell) 
  \qquad\text{for}\quad
  \za = 1 \, .
\end{eqnarray}
\end{subequations}%
Hence, the leading-order scaling of the auto-correlation function is given by:
\vskip 5pt \noindent
\begin{lem}\hskip -4pt{\rm \bf :}\label{thm:q.eq.1-scaling}
  For any $\ell > 0$ 
  the auto-correlation function, $\phi((1+\ell)\,m, m)$, 
  asymptotically scales as:
\begin{equation}\label{eq:q.eq.1-scaling}
  \phi((1+\ell)\,m, m) 
  \sim 
  \left\{
  \begin{array}{lll}
    \frac{2}{1-\za} \: 
    \left( \left( 1 + \ell \right)^{1-\za} - \frac{\za}{2-\za}  \right) \: m^{2-\za}
    & \quad\text{for }
    & 0 < \za < 2, \quad \za \neq 1 \, ,
    \\[2mm]
    \left( 4 + 2\, \ln( 1 + \ell ) \right) \: m
    & \quad\text{for }
    & \za = 1 \, ,
    \\[2mm]
    4\: \ln( m )
    & \quad\text{for }
    & \za = 2 \, ,
    \\[2mm]
    \text{const}
    & \quad\text{for }
    & \za > 2 \, .
  \end{array}
  \right . 
\end{equation}
\end{lem}
\vskip 5pt
\noindent
\proof 
The cases $\za \geq 2$ are obtained as in Lemma~\ref{thm:fixed-m-scaling}.

For $0 < \za < 2$ the leading-order contributions to all three sums in \Eq{intgrl corr.expr.} 
scale like $m^{2-\za}$.  
The case $\za=1$ is special, however, because the second sum takes a different prefactor, \Eq{2m.sum.k.Delta-q.eq.1B}, rather the one obtained in \Eq{2m.sum.k.Delta-q.eq.1A}.
For  $0 < \za < 2$ and $\za \neq 1$ we have
\begin{eqnarray*}
  \phi((1+\ell)\,m, m) 
  \sim
  \left( \frac{2\,\za}{2-\za} 
       + \frac{2\,\za}{1-\za} \; \left( (1+\ell)^{1-\za} - 1 \right) 
       + 2 \: (1+\ell)^{1-\za}
  \right) \;
  m^{2-\za} \, ,
\end{eqnarray*}
while for  $\za = 1$ we have
\begin{eqnarray*}
  \phi((1+\ell)\,m, m) 
  \sim
  \left( 2
       + 2 \: \ln(1+\ell)
       + 2 
  \right) \;
  m \, .
\end{eqnarray*}
The result indicated in \Eq{q.eq.1-scaling} is obtained after collecting terms.
\hfill
$\Box$
\vskip 8pt \noindent

\paragraph{Scaling for $q>1$. \ }

In this case \Eq{2m.sum.k.Delta-q} still applies, but $n^{1-\za}$ is the dominating term in the bracket 
for $\za > 1$, while
$m^{1-\za}$ is the dominating term in the bracket for $\za < 1$.
Moreover, the logarithm in the scaling provided in \Eq{2m.sum.k.Delta-q.eq.1B} now scales as
$\ln(n/m) = \ln(1+\ell\,m^{q-1}) \sim (q-1) \, \ln(m)$.
When we further observe that $n \sim \ell \, m^q$, this implies
\begin{eqnarray}\label{eq:2m.sum.k.Delta-q.grt.1}
  2\, m \sum\limits_{k=m+1}^{n} k \, \Delta_k(\za) 
  & \sim &
    \left\{
    \begin{array}{lll}
          \frac{2 \, \ell\, \za}{1-\za} \: m^{1+q\,(1-\za)} 
          & \quad\text{for }
          & 0 < \za < 1 \, ,
          \\[2mm]
          2 \, (q-1) \: m \: \ln(m)
          & \quad\text{for }
          & \za  = 1 \, ,
          \\[2mm]
          \frac{2 \, \za}{\za - 1} \: m^{2-\za} 
          & \quad\text{for }
          & \za > 1 \, .
    \end{array}
  \right .
\end{eqnarray}
Hence, the leading-order scaling of the auto-correlation function obeys the following:
\vskip 5pt \noindent
\begin{lem}\hskip -4pt{\rm \bf :}\label{thm:q.gtr.1-scaling}
  For  $q>1$, $\ell > 0$, and  $m \to \infty$ the auto-correlation function, 
  $\phi(m + \ell\, m^q, m)$ asymptotically scales as: 
\begin{equation}\label{eq:q.grt.1-scaling}
  \phi(m+\ell\,m^q, m) 
  \sim 
  \left\{
  \begin{array}{lll}
    \frac{2}{1-\za} \: \ell^{1-\za} \: m^{1+q\,(1-\za)}
    & \quad\text{for }
    & 0 < \za < 1 \, ,
    \\[2mm]
    2\, (q-1) \: m \: \ln( m )
    & \quad\text{for }
    & \za = 1 \, ,
    \\[2mm]
    \frac{2\, \za}{(2-\za)\, (\za-1)} \: m^{2-\za}
    & \quad\text{for }
    & 1 < \za < 2 \, ,
    \\[2mm]
    4\, \ln( m )
    & \quad\text{for }
    & \za = 2 \, ,
    \\[2mm]
    \text{const}
    & \quad\text{for }
    & \za > 2 \, .
  \end{array}
  \right . 
\end{equation}
\end{lem}
\vskip 5pt
\noindent
\proof 
The cases $\za \geq 2$ are obtained as in Lemma~\ref{thm:fixed-m-scaling}.

For $1 < \za < 2$ the leading-order contributions scale like $m^{2-\za}$.
They appear in the first and in the second sum on the right-hand-side of \Eq{corr.expr.}.
Collecting the corresponding terms in \Eqs{sumDeltaKsquareScaling} and \eq{2m.sum.k.Delta-q.grt.1}
we obtain
\begin{eqnarray*}
  \phi(m+\ell\,m^q, m)
  \sim
  \left( \frac{2\,\za}{2-\za} + \frac{2\,\za}{\za-1} \right) \: m^{2-\za}
  =
  \frac{2\,\za}{(2-\za)\,(\za-1)} \: m^{2-\za} \, .
\end{eqnarray*}

For $\za = 1$ the leading-order scaling contribution to the auto-correlation function is provided in \Eq{2m.sum.k.Delta-q.grt.1}.

For $0 < \za < 1$ the leading-order contributions scale like $m \, n^{1-\za} \sim \ell^{1-\za}\, m^{1+q\,(1-\za)}$.
Collecting these terms in \Eqs{sumDeltaScaling} and \eq{2m.sum.k.Delta-q.grt.1} provides
\begin{eqnarray*}
  \phi(m+\ell\,m^q, m)
  \sim
  \left( \frac{2\,\za}{1-\za} + 2 \right) \: \ell^{1-\za}\: m^{1+q\,(1-\za)}
  =
  \frac{2}{1-\za} \:  \ell^{1-\za} \: m^{1+q\,(1-\za)} \, .
\end{eqnarray*}
\hfill $\Box$

%%%%  ------------------------------------------------------------------------------------------   %%%%
%%%%  ------------------------------------------------------------------------------------------   %%%%
\section{Comparison of the SM with a L\'evy-Lorentz gas}
\label{sec:compareLW}

The LLg is a random walk in a one-dimensional random environment \cite{BFK00}, 
where a point particle moves ballistically (with velocity $\pm v)$ between static point scatterers.
At each scatterer the particle is either transmitted or reflected with probability $1/2$. 
The distance $r$ between two consecutive scatterers is a random variable 
drawn independently and identically from a L\'evy distribution with density:
\begin{equation}\label{eq:levydensity}
\lambda(r) = \zb r_0^\zb \frac{1}{r^{\zb+1}},\quad r\in [r_0, +\infty),
\end{equation}
where $\zb>0$, and $r_0$ is the characteristic length scale of the system.

The LLg shares basic similarities with the SM in that both systems deal with non-interacting particles 
and the initial condition plays an important role. 
On the other hand, the differences are evident: 
The LLg is a continuous-time stochastic system, 
while the slicer dynamics is  discrete-time and deterministic. 
In particular, the LLg dependence on the initial conditions is considerably more intricate than in the SM:
the LLg transport properties depend on whether a walker can start its trajectory away from the scatterers, 
called equilibrium initial condition, or must start exactly at a scatterer, called non-equilibrium 
initial condition \cite{BFK00,BCV10}.
The asymptotic behaviour of the moments is known for the LLg with non-equilibrium initial conditions.
Hence, we focus on this situation, and we show that for this setting
the SM provides insight into transport properties of the LLg.

%%%%  ------------------------------------------------------------------------------------------   %%%%
\subsection{Moments of the Displacement}\label{sec:Pmoment}

\citet{BFK00} calculated bounds for the mean-square displacement for equilibrium and non-equilibrium initial conditions.
Subsequently, \citet{BCV10} adopted some simplifying assumptions to find the asymptotic form 
for non-equilibrium conditions of all moments $\langle |r(t)|^{p}\rangle$ with $p>0$\,:
\begin{equation}\label{eq:MOMBUR}
  \langle |r(t)|^{p} \rangle
  \sim
  \left\{
    \begin{array}{lll}
      t^{\frac{p}{1+\zb}}                  & \quad\text{for  } & \zb<1,\ p<\zb    \, ,\\
      t^{\frac{p(1+\zb)-\zb^{2}}{1+\zb}}      & \quad\text{for  } & \zb<1,\ p>\zb    \, ,\\
      t^{\frac{p}{2}}                     & \quad\text{for  } & \zb>1,\ p<2\zb-1 \, ,\\
      t^{\frac{1}{2}+p-\zb}                & \quad\text{for  } & \zb>1,\ p>2\zb-1 \, . 
    \end{array}
  \right. 
\end{equation}
For the mean-square displacement, $p=2$, this result implies
\begin{equation}\label{eq:LLgMeanSquare}
  \langle r(t)^{2} \rangle 
  \sim 
  t^\zmsd 
  \quad\text{with}\quad 
  \zmsd 
  =
\left\{
    \begin{array}{lll}
      2 - \frac{\zb^{2}}{(1+\zb)}   & \quad\text{for  } & \zb < 1    \, ,\\
      \frac{5}{2}-\zb              & \quad\text{for  } & 1 \leq \zb < 3/2  \, ,\\
      1                            & \quad\text{for  } & 3/2 \leq \zb         \, .
    \end{array}
  \right. 
\end{equation}%
Unlike the SM case, that enjoys sub-diffusive transport for $\za > 1$, 
non-equilibrium initial conditions for the LLg only lead to 
super-diffusive ($0<\zb< 3/2$) or 
diffusive ($\zb \ge  3/2$) regimes: sub-diffusion is not expected.

\citet{Salari} observed that the moments of the SM in its super-diffusive regime ($0<\za < 1$) can be mapped to those of the LLg. 
They proved that all moments of the SM, \Eq{DeltaMoments}, scale like those conjectured and numerically validated for the LLg, \Eq{MOMBUR}, once the second moments do. 
This is the case if the following holds, \cf~\Eqs{MeanSquareDisplacement} and \eq{LLgMeanSquare}:
\begin{equation}\label{eq:llgpdf}
  \za 
  =
  \left\{
    \begin{array}{lll}
      \frac{\zb^{2}}{(1+\zb)}     & \quad\text{for  } & 0 < \zb \leq 1          \, ,\\[2mm]
      \zb - \frac{1}{2}          & \quad\text{for  } & 1 < \zb \leq \frac{3}{2}\, ,\\[2mm]
      1                          & \quad\text{for }  & \frac{3}{2} < \zb \, .
    \end{array}
  \right. 
\end{equation}
When adopting this mapping also all other moments of the SM agree with those of the LLg, \Eq{MOMBUR}.

This means that relation \Eq{llgpdf} makes the SM and the LLg asymptotically indistinguishable from 
the viewpoint of moments, provided the assumptions of \cite{BCV10} holds.
This equivalence is by no means trivial. 
In particular, the relation takes different functional forms in different parameters ranges, 
because the LLg has different scaling regimes for super-diffusive transport, 
while the SM has only one regime for all kinds of transport. 
We now explore whether the position-position auto-correlations of the two dynamics differ.
The correlations are calculated analytically for the SM.
This data will then be compared to numerical data for the LLg. 
For correlations in the LLg there are no analytic results such as those of \cite{BCV10} for the moments.

%%%%  ------------------------------------------------------------------------------------------   %%%%
\subsection[Numerical Implementation]{Numerical Implementation of the L\'evy-Lorentz Gas}

The non-equilibrium initial conditions for the LLg are implemented by starting each particle in the origin $x_0=0$,
where a scatterer is assumed to be present in all realisations of the scatterers distributed in the line $\R$.
Moreover, trajectories that return to the origin provide a minor contribution to the moments for super-diffusive transport.
For numerical tests, given the symmetry of the dynamics, we modify the original dynamics of the LLg,
placing a reflecting barrier at $x=0$ and giving an initial positive velocity to each LLg walker.
Thus, the resulting system, denoted \llgp, which we numerically verified to yield the
same results of the LLg for the position auto-correlation function, evolves in $\R^+_0$, similarly to
the SM with initial conditions in $(1/2, 1) \times \{0\}$,
that evolve in the half configuration space $\widehat{M}^+$.

More precisely, the setting is as follows. Let $(L_0,L_1,L_2,\ldots )$ be a sequence of i.i.d.\ random variables with density \Eq{levydensity}, and let $Y_{i+1}=Y_{i}+L_i,\, i=0,1,2,\ldots$,
with $Y_0\equiv 0$. Denote by ${\tt Y}$ a given realisation of the sequence
$(Y_0\equiv0, Y_1,Y_2,\ldots )$,  that represents one random
scatterers distribution in $\R^+_0$.
We introduce the discrete-time process that represents the \llgp at the scattering events.
Let $\omega=(\omega_0,\omega_1,\omega_2,\ldots )$ be a random walk on $\Z^+_0$
with the conditions
that $\omega_0\equiv 0$ and $\omega_n- \omega_{n-1},\, n=1,2,\ldots $ are i.i.d.\ dichotomic variables,
known as Rademacher random variables,
with $P(\omega_n- \omega_{n-1}=+1\, |\,  \omega_{n-1}\ne 0 )=P(\omega_n- \omega_{n-1}=-1\, |\,  \omega_{n-1}\ne 0 )=1/2$,
and  $P(\omega_n- \omega_{n-1}=+1\, |\,  \omega_{n-1}= 0 )=1$.
These conditions mean that the walk starts at $0$ and whenever it returns there,
it is reflected to the right.
Away from $0$, each walker follows a simple symmetric random walk.
Then, the process that represents the position of the moving particle at the scattering events
will be given by   ${\cal W}=(Y_{\omega_0}, Y_{\omega_1}, Y_{\omega_2},\ldots)$. From knowledge of ${\cal W}$,
the continuous-time position $r(t)$
of the corresponding moving particle of the \llgp\
can be unambiguously reconstructed,
because the velocity  between any two scattering events is constant.

The process $r(t)$ is affected by two sources of stochasticity:
the environment $\tt Y$ and the scattering $\omega$.
Hence, averages can be taken in two different fashions.
Let us denote by $\Eo$ the average w.r.t.\ the process $\omega$,
\ie\ the average over particles that can be identified with their scattering sequences
in a given realisation of the environment.
Analogously, let $\Ey$ denote the average over the random scatterers realisations.
Then, the average of $r^2(t)$ at fixed scatterers configuration $\tt Y$,
is denoted by $\Eo(r^2(t) | {\tt Y})$.
This is a random quantity because $\tt Y$ is random.
Averaging this quantity over the ensemble of scatterers yields the mean-square displacement of the \llgp:
\begin{equation}\label{eq:msdllgp}
\langle r^2(t) \rangle_\zb = \Ey [ \Eo(r^2(t) | {\tt Y}) ] \, .
\end{equation}
The subscript $\zb$ indicates that the distribution of scatterers, \Eq{levydensity}, depends on $\zb$.

This procedure has been implemented in a FORTRAN code by introducing a truncation in the sequence of
scattering events $\bar{\omega}=(\omega_0,\omega_1,\omega_2,\ldots, \omega_N )$ that corresponds to a time
$T=T(\bar{\omega},{\tt Y})$ at which the process $r(t)$ stops. The stopping time $T(\bar{\omega},{\tt Y})$
is random,
and typically large if the scatterers are placed at large distances from one another,
\ie\ for small $\zb$.
In contrast, for large $\zb$, the distances are on average approximately equal $r_0$.%
\footnote{More precisely, if $L$ is distributed according to \Eq{levydensity},
then $\E [L]=+\infty$ if $\zb \le 1$ and
$ \E [L]=\frac{\zb r_0}{\zb -1}$ if $\zb > 1$.
Moreover the variance is
$\frac{\zb\, r_0^2}{(\zb-2)(\zb-1)^2}$ for $\zb >2$ and $+\infty$ for $\zb\le 2$.
For $\zb \gg 1$ the expected distance is therefore $\E [L] \simeq r_0$ with relative deviations of the order of $\zb^{-1}$.}
Therefore, one expects typically smaller and smaller $T(\bar{\omega},{\tt Y})$
for larger and larger $\zb$,
with the risk of under-sampling the large-time behaviour of the \llgp\
in numerical estimates of statistical properties.
We do not present data with insufficient statistics.

Our choice of $r_0$ and $v$ in the  numerical simulations of the \llgp\ follows \citet{BCV10}.
We set the characteristic length $r_0$ to~$0.1$, 
and the velocity $v$  of the ballistic motion is always $1$.
The number of simulated scattering events is  $N=2.5 \cdot 10^6$.

We tested the code and explored the relation between the LLg and the \llgp,
by calculating the mean-square displacement of the \llgp,
in order to verify the power-law behaviour of the LLg, see \Eq{LLgMeanSquare}.
Table \ref{tab:tLLpLLpp} shows that our numerical results for the mean-square displacement
for the \llgp\ accurately reproduce the exponent given in \Eq{LLgMeanSquare} for the LLg,
at least for not too large values of $\zb$.
These results, with similar ones obtained  by comparing various position-position auto-correlation
functions of the two models, indicate the equivalence of the LLg and the \llgp,
at least at the level of the mean-square displacement and some correlation functions.
The slightly decreasing accuracy for increasing $\zb$, observed in Table \ref{tab:tLLpLLpp} and in the computation of correlations, can be attributed
to poorer statistics  of the numerical estimates, as suggested above.
Therefore, in the following we mainly focus on the cases with $\zb \lesssim 1$,
while more accurate data for larger $\zb$ will be presented in forthcoming work.

Finally, we observe that our simulations concern the \llgp\
because they are computationally more efficient than simulations of the LLg.
This can be heuristically understood by observing that,
at a fixed simulation length,
the \llgp\ dynamics produce trajectories
that typically reach larger distances from the origin, than those reached by trajectories of the LLg.
This provides better sampling for the long-time behaviour. 

%%  ------------------------------------------------------------------------------------------
\begin{table}
  \begin{center}
    \begin{tabular}{  c @{$\qquad\qquad$}  l @{$\qquad\qquad$}  c   p{5cm}}  
    \hline
    $\zb$ & $\zmsd$ &
      \\ 
          & \Eq{LLgMeanSquare} & fit to data 
      \\ 
   \hline
    $0.1$ & $1.99$ & $1.99$ %& ---
    \\     
     %\hline
    $0.3$ & $1.93$ & $1.93$ %& ---
    \\ 
    %\hline
    $0.5$ & $1.83$ & $1.82$ %& $8 \times 10^{-3}$
    \\
    %\hline
    $0.6$ & $1.77$ & $1.73$ %& $2 \times 10^{-2}$ 
    \\ 
    %\hline
    $0.8$ & $1.64$ & $1.63$ %& $7 \times 10^{-3}$ 
    \\
    %\hline
    $1.0$ & $1.50$ & $1.51$ %& $4 \times 10^{-3}$ 
    \\
    %\hline
    $1.3$ & $1.20$ & $1.18$ %& $1.5 \times 10^{-2}$ 
    \\
    %\hline
    $2.0$ & $1.00$ & $0.95$ %& $5 \times 10^{-2}$ 
    \\
    \hline
    \end{tabular}
    \caption{\label{tab:tLLpLLpp} 
      Comparison of numerical values for the scaling exponent $\zmsd$ 
      of the mean-square displacement in the \llgp\ (third column) 
      vs.~the prediction of \Eq{LLgMeanSquare} (second column).
      The numerical estimate of $\zg$ agrees with the expressions
      for the LLg derived and tested in \citet{BCV10}. 
    }
  \end{center}
\end{table}
%%  ------------------------------------------------------------------------------------------

%%%%  ------------------------------------------------------------------------------------------   %%%%
\subsection{Correlations of the \llgp}
\label{sec:CorrLL}

For $t,s \ge 0$, we define the position-position auto-correlation function for the \llgp\ as follows:
\begin{equation}\label{eq:corrlw}
\varphi(t,s)= \langle r(t) \, r(s) \rangle_\zb = \Ey [ \Eo(r(t)r(s) | {\tt Y}) ].
\end{equation}
We aim at comparing  the asymptotic behaviour of  $\varphi(t,s)$ 
with that of the  SM auto-correlation function  $\phi(n,m)$,  \Eq{autocorrelation}.  
Following the scaling adopted in Sec.~\ref{sec:sub31}, \ref{sec:sub32}, and \ref{sec:sub33} 
for the SM we consider three cases: 
\begin{itemize}
\item[1. ] $\varphi(t,s)$ for $t \to \infty$ at a fixed value of $s$.
\item[2. ] $\varphi(t+\tau, t)$ for $t \to \infty$ at a fixed value of $\tau$.
\item[3. ]  $\varphi(t+\ell \, t^q, t)$ for $t \to \infty$ at fixed $q$ and $\ell>0$.
\end{itemize}
Note that there is no free fit parameter in this comparison of the exponents, when one assumes 
the relation \Eq{llgpdf} between $\za$ and $\zb$. 
Here, we verify that $\za$ and $\zb$ obey \Eq{llgpdf} 
when the asymptotic scalings of the position-position auto-correlation functions of the SM and the \llgp\ match.

%%%%  ------------------------------------------------------------------------------------------
\subsubsection{Correlation $\langle r(t) \, r(s) \rangle_\zb$ with $s>0$ constant}

In Sec.~\ref{sec:sub31} we provided the scaling of the position-position
auto-correlation function for the SM, \Eq{fixed-m-scaling}, 
when one of its times is fixed.
For $0 < \za < 1$ we have:
\begin{equation}\label{eq:corrsd1}
  \langle \Delta x_n \, \Delta x_m \rangle_\za
  \sim 
  \frac{2 \, m}{1-\za}\, n^{1-\za},\quad \mbox{as}\,\, n\to \infty.
\end{equation}
Here and in the following we denote by $ \langle \cdot \rangle_\za$ the ensemble average of the trajectories of the SM with parameter $\za$. 
In analogy to the scaling, \Eq{corrsd1}, we propose the following
%
%% ----------------------------------------------------------------------------------------------------- %%
\vskip 5pt \noindent
\begin{conj}\hskip -4pt{\rm \bf a:}
The auto-correlation function of the \llgp\ asymptotically scales as the one of the SM.  
When the time $s$ is fixed, the auto-correlation function $\langle r(t) \, r(s) \rangle_\zb$ obeys:
\begin{subequations}
\begin{align}\label{eq:corrllgp1}
  \lim_{t \to \infty} 
  \frac{ \langle r(t) \, r(s) \rangle_\zb }{ \frac{ 2 \, s }{w_1}  \, t^{w_1}} 
  & =  C_1  \ne 0 \, ,
  \\[2mm]
  \mbox{with}\qquad w_1 = 1-\za(\zb) = \zg(\zb)-1 \, .
\end{align}
\end{subequations}
\end{conj}
%% ----------------------------------------------------------------------------------------------------- %%
%
%%  ------------------------------------------------------------------------------------------
\begin{figure}
{\hfill
\subfloat[\label{fig:corr__fix_s_2000} Varying $\zb$ for fixed $s=2000$.]
{\includegraphics[width=0.47\textwidth]{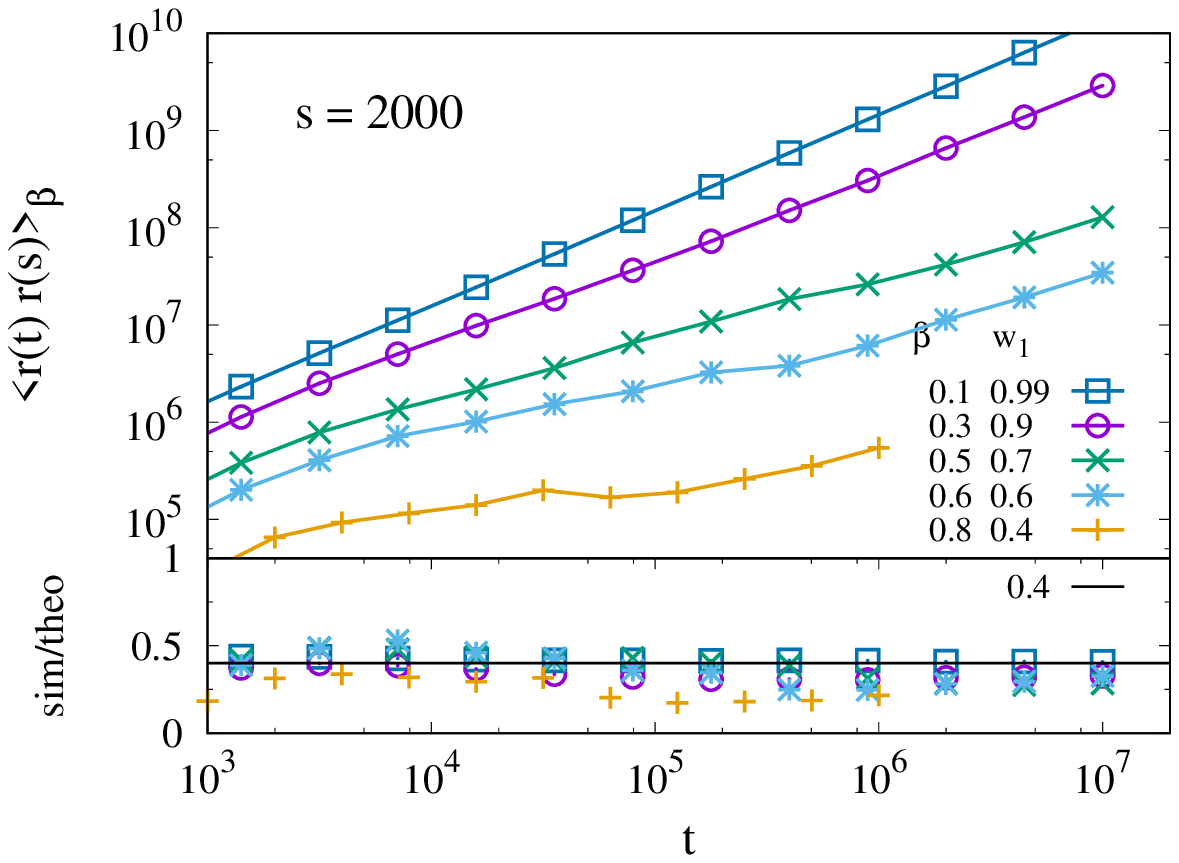}}
\hfill
\subfloat[\label{fig:corr__fix_s_b_01} Varying $s$ for fixed $\zb=0.1$.]
{\includegraphics[width=0.47\textwidth]{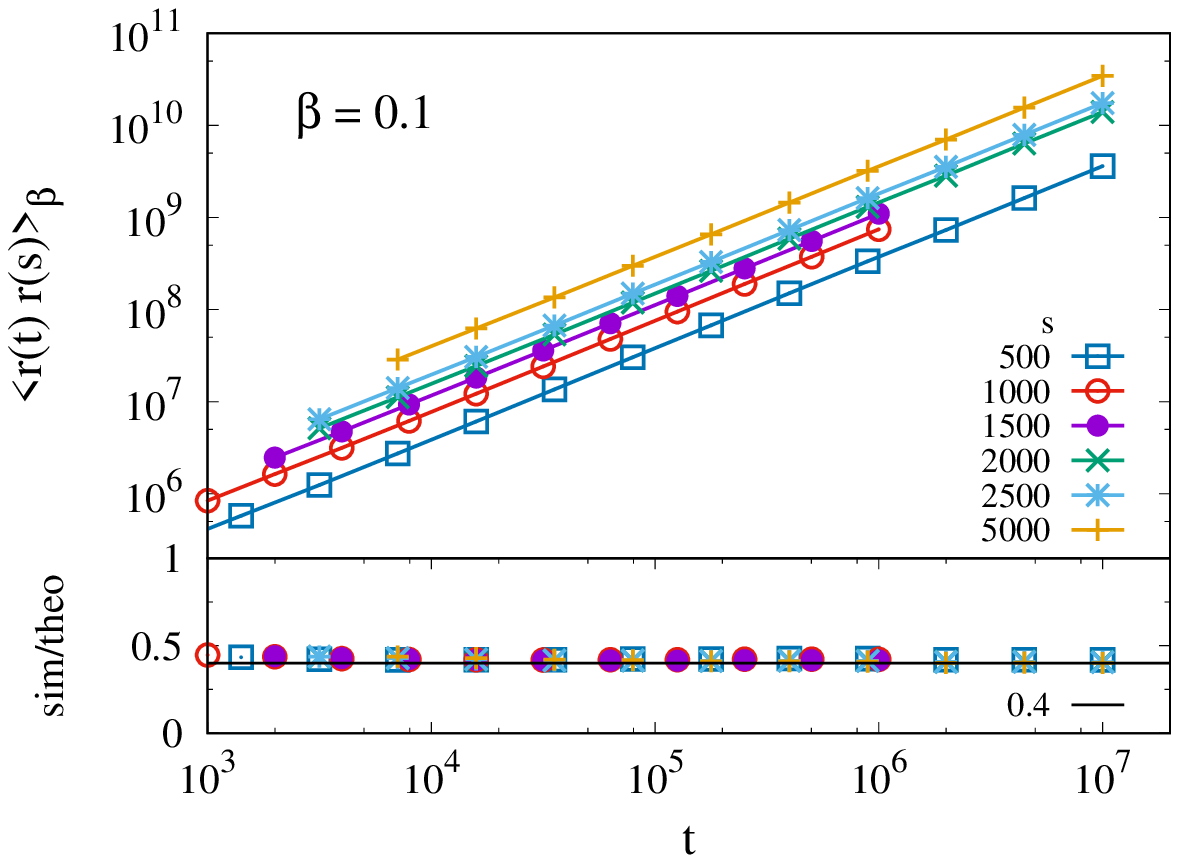}}
\hfill}
\caption[]{
  Log-log plots of the  correlation $\langle r(t) \, r(s) \rangle_\zb$ as a function of time $t$, 
  for different values of $\zb$ and of $s$. 
  The respective values for $\zb$ and $s$ are specified in the figure legends.
  In Figure~\ref{fig:corr__fix_s_2000} we also specify the values for the exponents $w_1$ 
  that provide the best fit to the data.
  The approach of the data towards the solid line in the bottom panel demonstrates 
  that \Eq{corrllgp1} provides a faithful asymptotic scaling, with $C_1=0.4$.
}
\end{figure}
%%  ---------------------------------------
\noindent
\evidence
The \llgp\ correlations $\langle r(t) r(s) \rangle_\zb$ have been computed for several values of $s$.
Numerical results for fixed $s=2000$ and different values of $\zb$ between $0.1$ and $0.8$ are shown 
in the upper panel of Figure~\ref{fig:corr__fix_s_2000}.
Moreover, in the upper panel of Figure~\ref{fig:corr__fix_s_b_01} we show 
data for $\zb=0.1$ and six values of $s$ in the range between $500$ and $5000$.

The respective lower panels show the time dependence of the ratio of \Eq{corrllgp1}, 
in order to test its asymptotic convergence.
For small $\zb$ and different $s$ this ratio provides a perfect data collapse
(Figure~\ref{fig:corr__fix_s_b_01}).
For larger $\zb$ the data collapse is still fair in view of the numerical accuracy of our data 
(Figure~\ref{fig:corr__fix_s_2000}).
Moreover, the scaling exponents $w_1$ adopted to achieve the collapse depend on $\zb$ and they are independent of~$s$.
The $\zb$-dependence agrees with the values $w_1 = 1-\za(\zb) = \zg(\zb)-1$ suggested by the SM
(\cf~the values for $\zg(\zb)$ provided in Table~\ref{tab:tLLpLLpp}).
Consequently, the SM provides a faithful description of the \llgp\ auto-correlation function,
both as far as the exponents and the the parameter-dependence of the prefactor is concerned.
\hfill
$\Box$
\vskip 8pt \noindent

%%%  --------------------------------------------------------------------------------------------
\subsubsection{Correlation $\langle r(t+\tau) \, r(t) \rangle_\zb$ with $\tau>0$ constant}

In Sec.~\ref{sec:sub32}, we provided the scaling of the auto-correlation function for the SM, \Eq{fixed-h-scaling}%\Eq{fixed-m-scaling}, 
when the difference $h$ between the times is fixed. % \Eq{fixed-h-scaling}.
For $0 < \za < 2$ and fixed $h$, one has
\begin{equation}\label{eq:corrsd2}
  \langle \Delta x_{m+h} \, \Delta x_{m} \rangle_\za  
  \sim \frac{4}{2-\za} m^{2-\za},\quad \mbox{as}\,\, m\to \infty \, .
\end{equation}
In analogy to this scaling we propose the following
%
%% ----------------------------------------------------------------------------------------------------- %%
\vskip 5pt \noindent
\addtocounter{thm}{-1}
\begin{conj}\hskip -4pt{\rm \bf b:}
The auto-correlation function of the \llgp\ asymptotically scales like the one of the SM.  
When the time lag $h$ is fixed, the correlation function $\langle r(t) \, r(t+\tau) \rangle_\zb$
obeys:
\begin{subequations}
\begin{align}\label{eq:corrllgp2}
  \lim_{t \to \infty} 
  \frac{ \langle r(t+\tau) \, r(t) \rangle_\zb }{ \frac{ 4 }{w_2}  \, t^{w_2}} 
  & =  C_2 \ne 0 \, ,
  \\[2mm]
  \text{with}\qquad w_2 = 2-\za(\zb) = \zg(\zb) \, .
\end{align}
\end{subequations}
\end{conj}
%% ----------------------------------------------------------------------------------------------------- %%
%%  ------------------------------------------------------------------------------------------
\begin{figure}
  {\hfill
    \subfloat[\label{fig:corr__fix_h_500} Varying $\zb$ for fixed $\tau=500$.]
    {\includegraphics[width=0.47\textwidth]{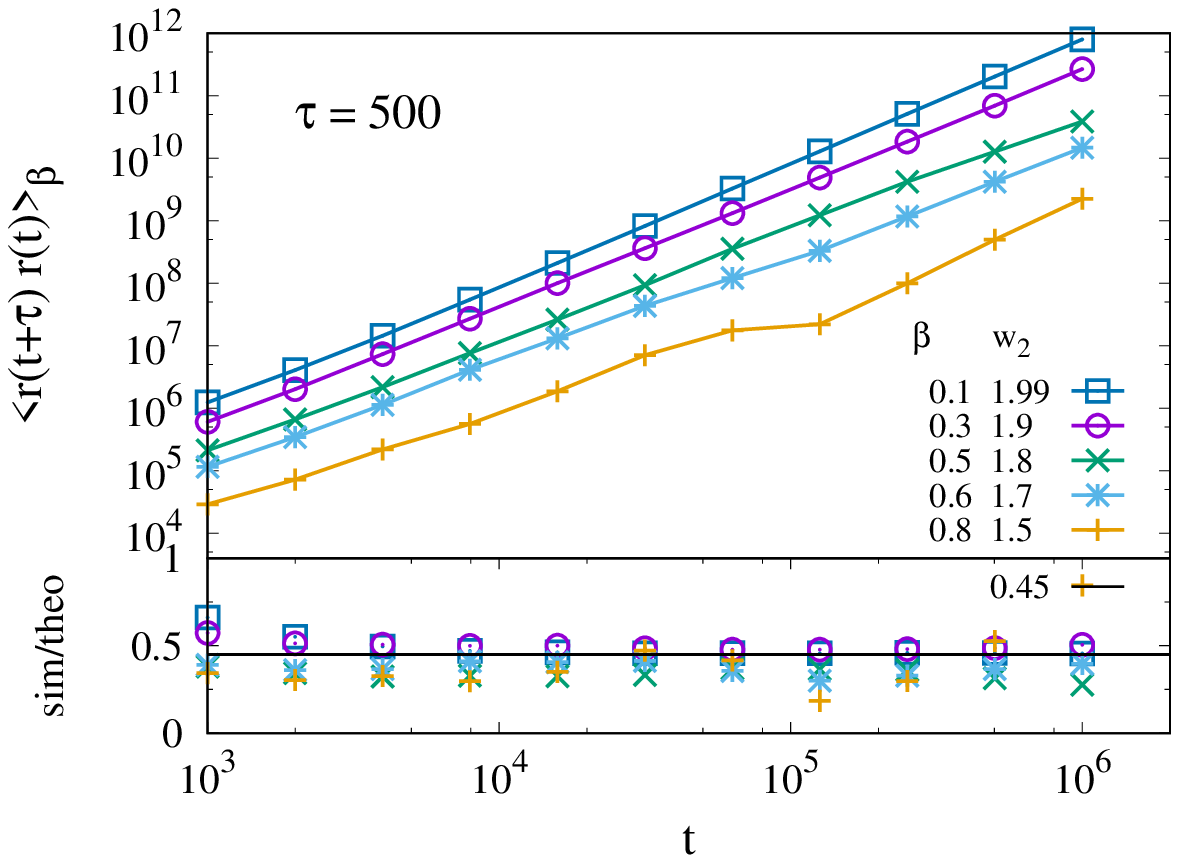}}
    \hfill
    \subfloat[\label{fig:corr__fix_h_b_01} Varying $h$ for fixed $\zb=0.1$.]
    {\includegraphics[width=0.47\textwidth]{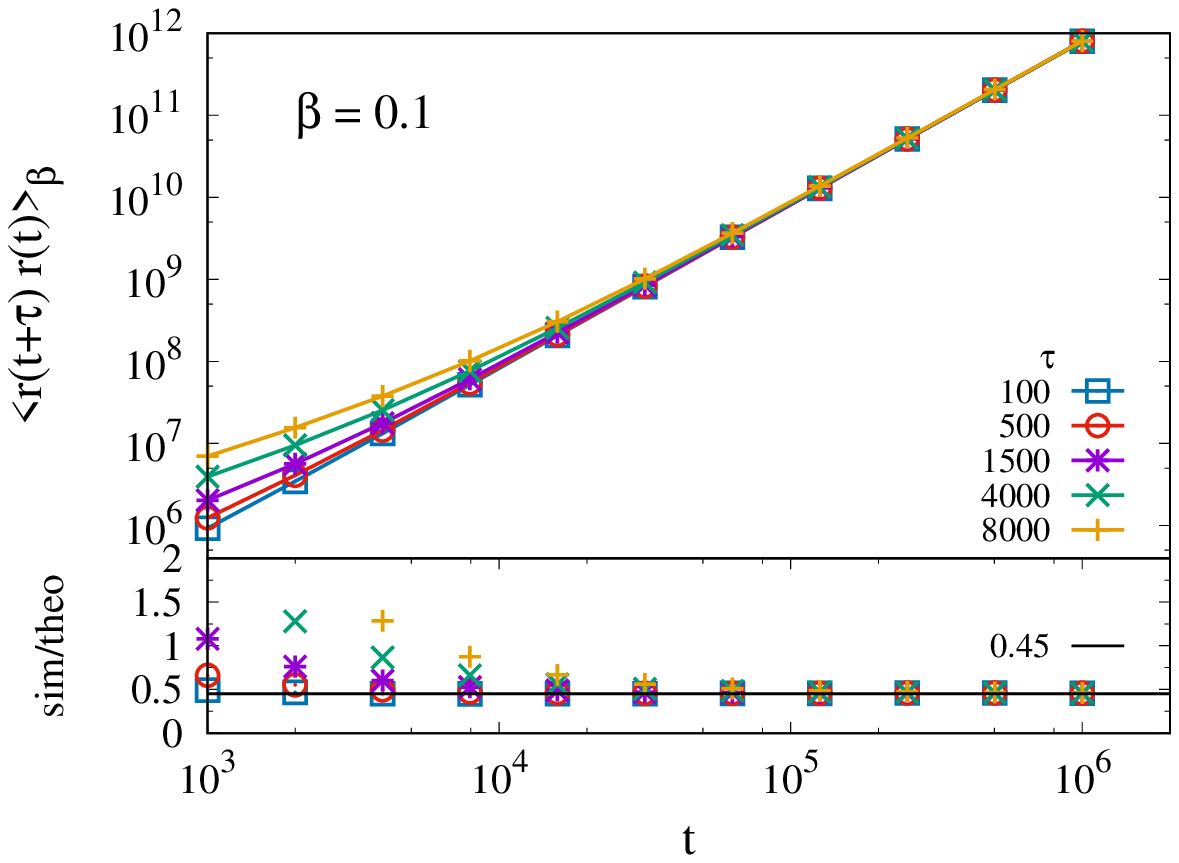}}
    \hfill
  }
  \caption[]{\label{fig:corr__fix_h}
    Log-log plots of the  correlation $\langle r(t+\tau)\, r(t) \rangle_\zb$  as a function of time $t$ 
    for various values of $\zb$ and of $\tau$. 
    The respective values for $\zb$ and $\tau$ are specified in the figure legends.
    Figure~\ref{fig:corr__fix_h_500} also specifies the values for the exponents $w_2$ 
    that provide the best fit to the data.
    The approach of the data towards the solid line in the bottom panel demonstrates that \Eq{corrllgp2} 
    provides a faithful asymptotic scaling, with $C_2=0.45$.
  }
\end{figure}
%%  ------------------------------------------------------------------------------------------
\vskip 5pt
\noindent
\evidence
In Figure~\ref{fig:corr__fix_h} we show numerical data for 
(a) a fixed value $\tau=500$ and $\zb$ in the range between $0.1$ and $0.8$, and
(b) a fixed value $\zb=0.1$ and $\tau$ in the range between $100$ and $8000$.
The lower panels show the ratio of the numerical data and the theoretical prediction, \Eq{corrllgp2}.
The curves  are not globally linear in the log-log plot.
However, they approach a power law for sufficiently  large values of $t$, 
and in that range they nicely follow the asymptotic scaling, \Eq{corrllgp2},
with $C_2=0.45$.
The coefficient  and the exponent of the asymptotic law are independent of $\tau$ 
and the dependence of $w_2$ faithfully agrees with the expected value $2-\za(\zb)=\zg(\zb)$,
as provided in Table~\ref{tab:tLLpLLpp}.
\hfill $\Box$
\vskip 8pt \noindent

%%%%  ------------------------------------------------------------------------------------------ 
\subsubsection{Correlation $\langle r(t+\ell\, t^q) \, r(t) \rangle_\zb$ with $\ell=1$ and $0<q<1$ constant}

In Section~\ref{sec:sub33}, we derived the auto-correlation for the SM, \Eq{fixed-m-scaling}.
For $0<q<1$ and $0<\za < 1$, one has:
\begin{equation}\label{eq:corrqq1}
  \langle  \Delta x_{m+m^{q}} \, \Delta x_m \rangle_\za 
  \sim 
  \frac{4}{2-\za} \: m^{2-\za},
  \quad \mbox{as}\, \,  m\to \infty \, .
\end{equation}
In analogy to this scaling, we propose the following
%
%% ----------------------------------------------------------------------------------------------------- %%
\vskip 5pt \noindent
\addtocounter{thm}{-1}
\begin{conj}\hskip -4pt{\rm \bf c:}
The auto-correlation function of the \llgp\ asymptotically scales like the SM.  
For the time  lag $\ell t^q$ with $0<q<1$ between its two times,
the auto-correlation function $\langle r(t) \, r(t+\ell\,t^q) \rangle_\zb$ obeys:
\begin{subequations}
\begin{align}\label{eq:corrllgp3}
  \lim_{t \to \infty} 
  \frac{ \langle r(t+\ell\,t^q) \, r(t) \rangle_\zb }{ \frac{ 4 }{w_3}  \, t^{w_3}} 
  & =  C_3 \ne 0 \, ,
  \\[2mm]
  \text{with}\qquad w_3 = 2-\za(\zb) = \zg(\zb) \, .
\end{align}
\end{subequations}
\end{conj}
%% ----------------------------------------------------------------------------------------------------- %%
%
%%  ------------------------------------------------------------------------------------------
\begin{figure}
{\hfill
\subfloat[\label{fig:corr__fix_q_07} Varying $\zb$ for fixed $q=0.7$.]
         {\includegraphics[width=0.47\textwidth]{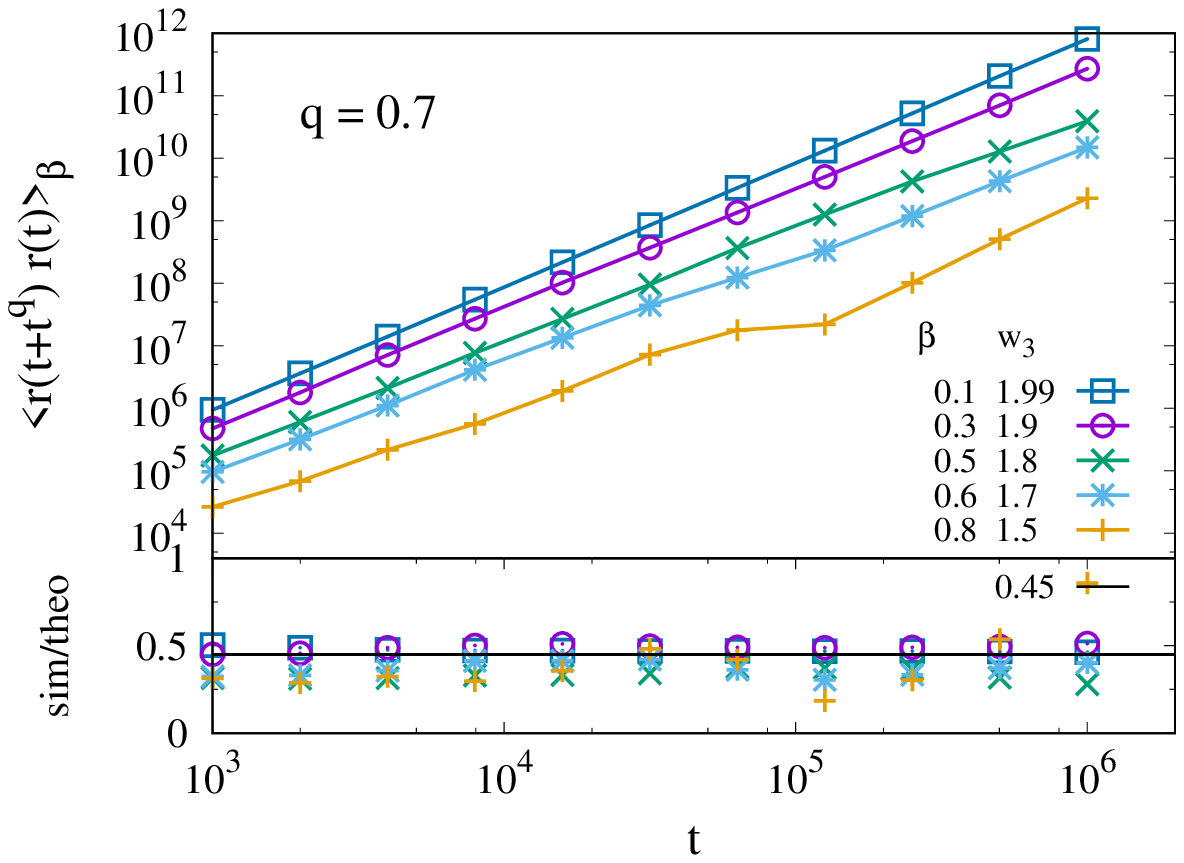}}
\hfill
\subfloat[\label{fig:corr__fix_q_b_01} Varying $q$ for fixed $\zb=0.1$.]
         {\includegraphics[width=0.47\textwidth]{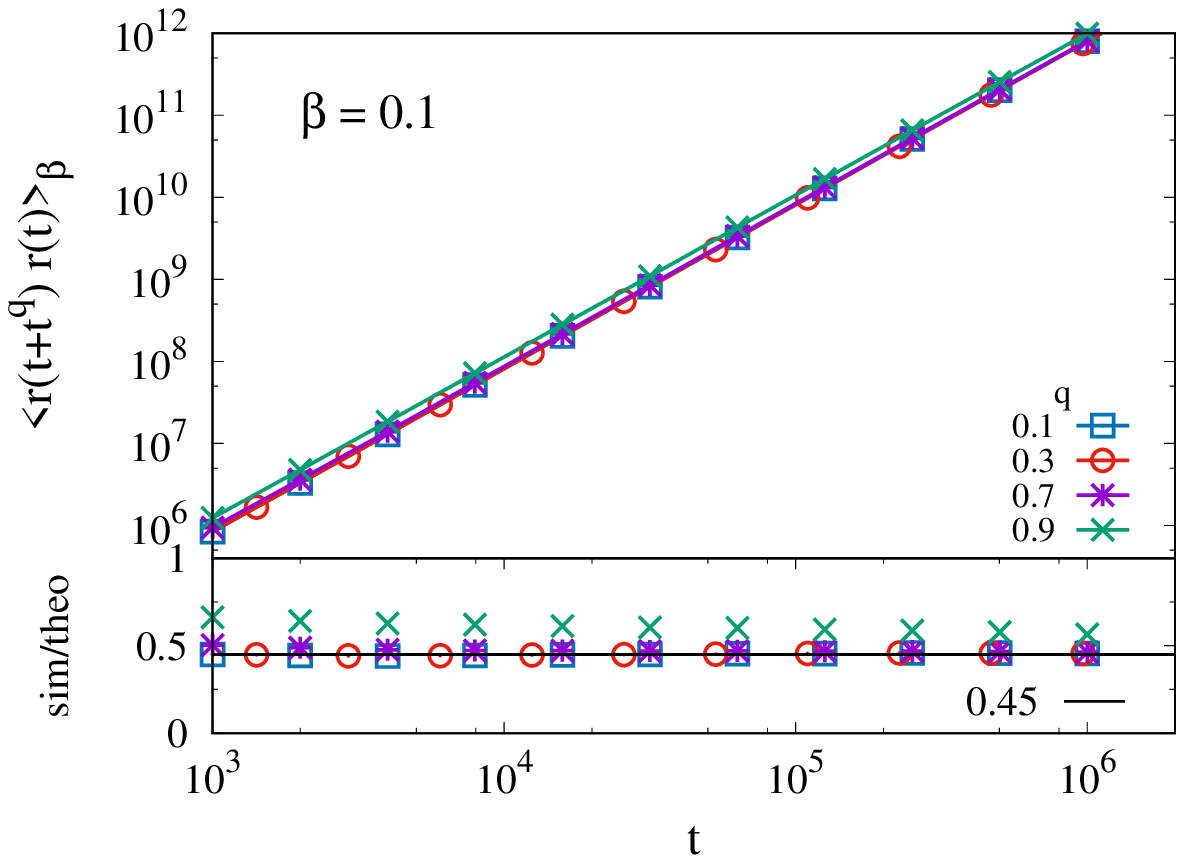}}
\hfill
}
\caption[]{\label{fig:corr__fix_q}
  Log-log plot of the correlation $\langle r(t)r( t+ t^q ) \rangle_\zb$  as a function of time $t$ 
for various values of $\zb$ and of $q$. 
  The respective values for $\zb$ and $q$ are specified in the figure legends.
  Figure~\ref{fig:corr__fix_q_07} also specifies the values for the exponents $w_3$ that provide the best fit to the data.
  Also in this case the dependence of $w_3$ agrees faithfully with the expected value $\zg(\zb)$
that has been provided in Table~\ref{tab:tLLpLLpp}.
  Further, the approach of the data towards the solid line in the bottom panel demonstrates again that \Eq{corrllgp3} provides a faithful asymptotic scaling, with $C_3=0.45$.
}
\end{figure}
%%  ------------------------------------------------------------------------------------------
\vskip 5pt
\noindent
\evidence
In Figure~\ref{fig:corr__fix_q}, we show numerical data for 
(a) a fixed value $q=0.7$ and $\zb$ in the range between $0.1$ and $0.8$, and
(b) a fixed value $\zb=0.1$ and $q$ in the range between $0.1$ and $0.9$.
The lower panels show the ratio of the numerical data and expected scaling, \Eq{corrllgp3}. 
Also in this case there is an excellent agreement between the data and the proposed asymptotic scaling.
\hfill $\Box$
\vskip 8pt \noindent
%%%%  ------------------------------------------------------------------------------------------   %%%%
%%%%  ------------------------------------------------------------------------------------------   
\section{Discussion}
\label{sec:discussion}
The investigation of the relation between the SM and the LLg started in \citet{Salari} with the demonstration of
the equivalence of the scalings of the time-dependent moments of the displacement.
Because it is well known
that moments do not sufficiently characterise transport processes \cite{Sokolov},
we have extended that study here to position-position auto-correlation functions.
We analytically computed the position-position auto-correlation function $\phi(n,m)$ of the SM, 
and we derived the asymptotic behaviour of this function in several cases 
corresponding to different relations between the times $m$ and $n$. 
Then, we numerically estimated the position-position auto-correlation function of the \llgp, 
in order to estimate its asymptotic behaviour.
The moments of displacement and the position-position auto-correlation functions of the \llgp\ agree with those of the LLg.
For the LLg there are theoretical results for the moments of displacement \citep{BCV10},
and they can be matched with the findings for the SM (Table~\ref{tab:tLLpLLpp} and \cite{Salari}).
In contrast, there are no analytical results available for the time dependent position-position auto-correlation function.
Time correlations in anomalous transport constitute by and large an open problem.  

Our numerical results indicate that there also is an equivalence of
the asymptotic scalings of the position-position auto-correlation
functions of the SM and the \llgp.  As established in
\citet{Salari} for the equivalence of moments, the agreement is
based on the matching of the transport exponent, $\zg$.
Hence, the parameters $\za$ and $\zb$ obey the relation \Eq{llgpdf}.
No further parameters are adjusted to also achieve the matching
  of the auto-correlation function.\footnote{Actually, we did not take the parameters suggested by the
    relation (\ref{eq:llgpdf}) for the equivalence of the moments, in
    order to find the data collapse for the correlations.
    On the contrary, we looked for the parameters that provide the
    best data collapse for the correlations, and we found that their
    values are indeed with good accuracy those given by \Eq{llgpdf}.
  }
As $\zb$ increases, the agreement between the numerical data and the proposed expressions for the 
asymptotic scaling of the auto-correlation function becomes less convincing.
Presently, it is not clear whether the correspondence only holds for small values of $\zb$, 
or whether the emerging discrepancies are due to the increasing difficulty of obtaining good 
statistics with growing $\beta$.
This issue goes beyond the scope of the present work. 
It will be investigated in a future paper.

We emphasise that the agreement of the moments and the time-dependent auto-correlation for the displacement hold
in spite of the fact that the SM and the \llgp\ exhibit entirely different dynamics.
Intuition on the properties of the SM and of the \llgp\ can be obtained 
by observing the relative motion of two points.
For the SM, take $\hX=(x,0)$ and $\hY=(y,0)$ in $\widehat{M}$.
There are two possible cases: either there exists an interval 
$(\ell^{+}_{j-1}(\alpha), \ell^{+}_{j}(\alpha)]$ such that $x,y\in (\ell^{+}_{j-1}(\alpha), \ell^{+}_{j}(\alpha)]$, 
or such an interval does not exist. 
\begin{enumerate}
\item 
When the interval exists the coarse-grained trajectories of $\hX$ and $\hY$, 
namely $x(n):=\pi_{\Z}(S^n_\alpha(\hX))$ and $y(n):=\pi_{\Z}(S^n_\alpha(\hY))$, 
coincide for all times $n$: 
particles sharing this property never separate.
For  all times they remain at a the initial distance from each other.
\item 
Otherwise, take 
$x \in (\ell^{+}_{\overline{m}_\alpha(x)-1}(\alpha), \ell^{+}_{\overline{m}_\alpha(x)}(\alpha)]$
and
$y \in (\ell^{+}_{\overline{m}_\alpha(y)-1}(\alpha), \ell^{+}_{\overline{m}_\alpha(y)}(\alpha)]$,
with
$(\ell^{+}_{\overline{m}_\alpha(x)-1}(\alpha), \ell^{+}_{\overline{m}_\alpha(x)}(\alpha)]
\cap
(\ell^{+}_{\overline{m}_\alpha(y)-1}(\alpha), \ell^{+}_{\overline{m}_\alpha(y)}(\alpha)] = \varnothing$ and 
$x<y$, which implies $\overline{m}_\alpha(x) < \overline{m}_\alpha(y)$.
This means that the two points have the same coarse grained trajectory up to time $\overline{m}_\alpha(x)$, when 
$\hX$ enters its periodic orbit, while $\hY$ continues its ballistic motion up to time $\overline{m}_\alpha(y)$. 
At times larger than $\overline{m}_\alpha(y)$,
the distance between the two particles equals either $\overline{m}_\alpha(y) - \overline{m}_\alpha(x)$ or 
$\overline{m}_\alpha(y) - \overline{m}_\alpha(x) \pm 1$. 
For all points with $x,y \in (0,1)$, the distance becomes periodic after a finite initial transient.
\end{enumerate}
Consequently, any function of any finite number of points, evaluated along a trajectory of the SM, 
turns periodic in a finite time. 
This situation is totally different from that of the \llgp, whose
nature renders the distance between any two particles, 
hence any function of a finite number of positions, stochastic.

We conclude that position-position auto-correlations do not distinguish the SM and the \llgp.
This equivalence can be used to indirectly investigate some of the elusive properties of the \llgp.
Given the non-physical features of the SM this may appear puzzling. 
However, from the perspective of statistical mechanics 
one could also argue that it is not surprising.
After all, in statistical mechanics the details of the microscopic dynamics of large systems
usually do not strongly affect the behaviour of physically relevant macroscopic quantities.
The latter can thus agree even for systems with vastly different microscopic dynamics.
This observation lies at the heart of the success of highly idealized models in describing 
complex phenomena; even simple models may capture the essential ingredients 
determining the behaviour of a selected and limited number of observables. 
Theoretical models for critical phenomena 
and universality constitute examples of this fact \cite{Sim93,Gal99,Kad00,CRV14}.
However, in general, one does not know how far equivalence can go, and which properties
it may concern, especially for far-from-equilibrium transient dynamics with
anomalous transport behaviour.
For instance, even for thermodynamic particle systems the Local Thermodynamic Equilibrium condition, 
required for the existence of the thermodynamic fields, is quite a sophisticated property 
whose underpinning requires a long sequence of microscopic conditions, 
as expressed by \citet{Spohn}: 
{\em ``The propagation of local equilibrium in time, if true, 
is a deep and highly non-obvious property of a system of many particles governed 
by Newton equations of motion''}.
For the position auto-correlations of the SM and the LLg or the \llgp
a direct investigation was therefore required.
The conclusion is that the SM can be used to indirectly investigate the \llgp:
agreement of transport exponents
implies matching of the two-point auto-correlation functions.
Thus, the transport exponent might be a kind of counterpart of critical exponents,
suitable for the characterisation of transport phenomena:
analogously to critical exponents,
they afford a coarse but equally useful description of the systems at hand. 

%%%%  ------------------------------------------------------------------------------------------   %%%%
\section*{Acknowledgements}

The authors are grateful to the computational resources provided by HPC$@$POLITO, 
a project of Academic Computing within the Department of Control and Computer Engineering 
at the Politecnico di Torino (\url{http://hpc.polito.it}). 
C.G. acknowledges financial support from ``Fondo di Ateneo per la Ricerca 2016'' 
under the project ``Sistemi stocastici e deterministici su strutture spaziali discrete, grafi e loro propriet\ah strutturali'',  Universit\ah di Modena e Reggio Emilia. 
J.V. is grateful for the appointment as a  Distinguished Visiting Professor 
at the   Department of Mathematical Sciences  of the  Politecnico  di Torino.

%%%%  ------------------------------------------------------------------------------------------   %%%%
\appendix
\section{Contributions to the Slicer Correlation Function}
\label{app:SumScaling}

In this Appendix we compute the asymptotics of the 
sums defined in \Eqs{sumDeltaKsquareScaling} and \eq{2m.sum.k.Delta-fix.m}.
By Taylor expansion we have:
\begin{subequations}
\begin{equation}\label{eq:slicereq12}
 \Delta_k(\za)
 = \ell^+_k(\za) - \ell^+_{k-1}(\za)
 = \frac{\za}{k^{\za+1}} \: 
   \left(1 - \tilde{c}(\za) \, \frac{1}{k} + O\left(\frac{1}{k^2}\right)\right) \, ,
\end{equation}
where
\begin{equation}
  \tilde{c}(\za) 
  = (1+\za) \: \left( 2^{\frac{1}{\za}}-\frac{1}{2}\right)>0 
  \qquad\text{with}\quad \za>0 \, .
\end{equation}
\end{subequations}
Then, the sum in \Eq{sumDeltaKsquareScaling} can be written as:
\begin{subequations}
\begin{equation}\label{eq:slicereq13}
  \sum\limits_{k=1}^{n-1} k^2 \, \Delta_k(\za) 
  = \sum\limits_{k=1}^{n-1} \frac{\za}{k^{\za-1}} \: \left(1-f(k)\right), 
\end{equation}
where
\begin{equation}\label{eq:slicereq14}
  f(k)
  :=
  \tilde{c}(\za) \, \frac{1}{k} + O\left(\frac{1}{k^2}\right) \, .
\end{equation}
\end{subequations}
The previous equation implies the existence of an integer $n_0$ such that:
\begin{eqnarray}\label{eq:slicereq15}
  \frac 1 2 \, \tilde{c}(\za) \, k^{-1} 
  < f(k) 
  < \frac 3 2 \, \tilde{c}(\za) \, k^{-1}
  \qquad\text{for}\quad k > n_0 \, .
\end{eqnarray}
Then, for $m>n_0$ we have:
\begin{subequations}
\begin{equation}\label{eq:slicereq16}
  \sum\limits_{k=1}^{n-1}    k^2 \Delta_k(\za) 
  = \sum\limits_{k=1}^{n-1}  \frac{\za}{k^{\za-1}} 
  - \sum\limits_{k=1}^{n_0}  \frac{\za \, f(k)}{k^{\za-1}} 
  - \sum\limits_{k=n_0}^{n-1} \frac{\za \, f(k)}{k^{\za-1}},
\end{equation}
where
\begin{equation}\label{eq:slicereq17}
  \frac 1 2 \tilde{c}(\za) \sum\limits_{k=n_0}^{n-1} \frac{1}{k^{\za}} 
  \leq \sum\limits_{k=n_0}^{n-1}  \frac{f(k)}{k^{\za-1}} 
  \leq \frac 3 2 \, \tilde{c}(\za) \sum\limits_{k=n_0}^{n-1} \frac{1}{k^{\za}} \, .
\end{equation}
\end{subequations}
Therefore, the last sum in \Eq{slicereq16} is of the order of $\sum_{k=n_0}^{n-1} k^{-\za}$, 
for $m\rightarrow\infty$. 
We evaluate the scaling of the two other terms based on the Euler-Maclaurin sum formula:
%
%% ----------------------------------------------------------------------------------------------------- %%
\begin{lem}\textrm{\bf (Euler-Maclaurin sum formula) }
For a smooth function $g(x)$, the full asymptotic behaviour of
\begin{equation}\label{eq:slicereq18}
G(n) = \sum\limits_{k=0}^n g(k),
\end{equation}
is given by
\begin{equation}\label{eq:eulerp}
  G(n)
  \sim  \frac{1}{2} g(n) 
  +     \int\limits_{0}^{n} g(t) \, dt 
  + C + \sum_{j=1}^{\infty} (-1)^{j+1} \frac{B_{j+1}}{(j+1)!} g^{(j)}(n)
  \qquad\text{as}\quad n \to \infty \, .
\end{equation}
Here $C$ is a constant depending on $g$, and $B_{j}$ are the Bernoulli numbers.

Specifically, for $g(k)=k^{p}$ and $p\ne -1$ one has
\begin{subequations}
\begin{equation}\label{eq:eulerS}
  \sum_{k=0}^{m} k^{p} 
  \sim \frac{m^{p+1}}{p+1} 
   +   \frac{1}{2} m^{p}  
   + C 
   + \sum_{j=1}^{\infty} \, (-1)^{j+1} \frac{B_{j+1}}{(j+1)!} \; \prod_{\ell=0}^{j-1} (p-\ell) \, m^{p-j} 
  \qquad\text{as}\quad  m\to \infty
\end{equation}
while $p=-1$ entails:
\begin{equation}\label{eq:eulerLog}
  \sum_{k=1}^{m} k^{-1} \sim \ln m + C + \frac{1}{2\,m} -\frac{B_{2}}{2\,m^{2}}-\frac{B_{4}}{4\,m^{4}}- \dots
  \qquad\text{as}\quad  m \to \infty \, .
\end{equation}
\end{subequations}
\end{lem}
%% ----------------------------------------------------------------------------------------------------- %%
\proof
See for instance \citet{BO}.
\hfill $\Box$
\vskip 5pt \noindent
%% ----------------------------------------------------------------------------------------------------- %%
\noindent
For $\za > 2$ \Eq{eulerS} entails that the sum \Eq{slicereq13} converges to a finite value,
as reported in \Eq{sumDeltaKsquareScaling}.

For $\za = 2$ \Eq{eulerLog} provides 
\begin{subequations}
\begin{equation}\label{eq:appKsquareScalingLog}
  \sum_{k=1}^{n-1} k^{2} \, \Delta_{k}(\za) 
  \sim 2 \za \; \ln n 
  \qquad\text{for}\quad \za = 2 \, , 
\end{equation}  
\ie~the logarithmic scaling reported in \Eq{sumDeltaKsquareScaling}.

For $0<\za < 2$ we have that the two sums depending on $m$ in \Eq{slicereq16} diverge:
\begin{align*}
  \sum_{k=1}^{m} \frac{\za}{k^{\za-1}} \sim \frac{\za}{2-\za} m^{2-\za} \, ,
  \qquad
  \sum_{k=m_{0}}^{m} \frac{\za f(k)}{k^{\za-1}}=O(m^{1-\za}) 
  \qquad\text{as}\quad  m\to \infty \, ,
\end{align*}
where the second sum is estimated using \Eq{slicereq17}. From \Eq{slicereq16}, we then obtain
\begin{equation}\label{eq:appKsquareScaling}
  \sum_{k=1}^{n-1} k^{2} \Delta_{k}(\za) 
  \sim \frac{\za}{2-\za} \; n^{2-\za} 
  \qquad\text{for}\quad 0<\za < 2 \, .
\end{equation}
\end{subequations}
This concludes the derivation of \Eq{sumDeltaKsquareScaling}.

\EQ{2m.sum.k.Delta-fix.m} can be evaluated by the same line of argumentation.
Using \Eq{slicereq12}, the 
sum in \Eq{2m.sum.k.Delta-fix.m} yields:
\begin{equation}
 \sum_{k=1}^{n} k \: \Delta_{k}(\za) 
  =
  \frac{\za}{1+\za} \; 
 \sum_{k=1}^{n}   
      \frac{1+\za}{k^{\za}} \:
      \left( 1 - f(k) \right)\,.
\end{equation}
This is the same expression as \Eq{slicereq13}, 
except for the constant factor in front of the sum 
and substituting $\za - 1 \to \za$ in the sum.
Consequently, according to \Eq{eulerS} the sum takes a finite value for $\za > 1$.
Moreover, introducing the substitutions into \Eq{appKsquareScaling} yields
\begin{equation}\label{eq:appKlinScaling}
  \sum_{k=1}^{n} k \: \Delta_{k}(\za) 
  \sim \frac{\za}{1+\za} \; \frac{1+\za}{1-\za} \; n^{1-\za} 
  = \frac{\za}{1-\za} \; n^{1-\za}
  \qquad\text{for}\quad 0<\za < 1 \, ,
\end{equation}
which is the non-trivial scaling reported in \Eq{2m.sum.k.Delta-fix.m}.
Analogously, the logarithmic scaling in \Eq{2m.sum.k.Delta-fix.m} is obtained from \Eq{appKsquareScalingLog}, where the right-hand-side must be evaluated for $\za=1$ due to the substitution.
This concludes the derivation of \Eq{2m.sum.k.Delta-fix.m}.

 %\bibliographystyle{plainnat}
 %\bibliography{./anomalous}

\end{document}